\documentclass[apj]{emulateapj}
\def\chandra    {\emph{Chandra}}
\def\xmm        {\emph{XMM}}
\def\vla        {VLA}

\def\gmrt {GMRT}

\def\lax{\lesssim}

\shorttitle{Mapping particle acceleration in RX\,J1720.1+2638}
\shortauthors{S.~Giacintucci et al.}

\begin{document}

\setlength{\pdfpageheight}{\paperheight}
\setlength{\pdfpagewidth}{\paperwidth}


\title{Mapping the particle acceleration in the cool core of the galaxy cluster RXJ1720.1+2638}

\author{S. Giacintucci\altaffilmark{1,2}, 
M. Markevitch\altaffilmark{3,2}, 
G. Brunetti\altaffilmark{4},
J.~A. ZuHone\altaffilmark{3},
T. Venturi\altaffilmark{4},
P. Mazzotta\altaffilmark{5},
H. Bourdin\altaffilmark{5}
}

\altaffiltext{1}{Department of Astronomy, University of Maryland,
  College Park, MD 20742, USA; simona@astro.umd.edu}
\altaffiltext{2}{Joint Space-Science Institute, University of Maryland, College Park,
MD, 20742-2421, USA}
\altaffiltext{3}{Astrophysics Science Division, NASA/Goddard Space
  Flight Center, Greenbelt, MD 20771, USA}
\altaffiltext{4}{INAF - Istituto di Radioastronomia, via Gobetti 101, I-40129 Bologna,
Italy}
\altaffiltext{5}{Dipartimento di Fisica, Universit\'a di Roma Tor
Vergata, Via della Ricerca Scientifica 1, I-00133, Rome, Italy}

\date{Received 03 - 12 - 2014; accepted 09 - 10 - 2014}

\begin{abstract}
We present new deep, high-resolution radio images of the diffuse minihalo in 
the cool core of the galaxy cluster RX\,J1720.1+2638. The images have 
been obtained with the Giant Metrewave Radio Telescope at 317, 617
and 1280 MHz and with the Very Large Array at 1.5, 4.9 and 8.4 GHz, 
with angular resolutions ranging from $1^{\prime\prime}$ to
$10^{\prime\prime}$. This represents the best radio spectral and imaging
dataset for any minihalo. Most of the radio flux of the minihalo
arises from a bright central component with a maximum radius of $\sim
80$ kpc. A fainter tail of emission extends out from the central component 
to form a spiral-shaped structure with a length of $\sim 230$ kpc,
seen at frequencies 1.5 GHz and below. We find indication of a
possible steepening of the total radio spectrum of the minihalo at high frequencies. 
Furthermore, a spectral index image shows that the spectrum of the diffuse emission 
steepens with the increasing distance along the tail. A striking spatial 
correlation is observed between the minihalo emission and two cold fronts 
visible in the {\em Chandra}\/ X-ray image of this cool core. These cold 
fronts confine the minihalo, as also seen in numerical simulations of 
minihalo formation by sloshing-induced turbulence. All these observations 
favor the hypothesis that the radio emitting electrons in cluster cool 
cores are produced by turbulent reacceleration. 
\end{abstract}

\keywords{galaxies: clusters: general -- galaxies: clusters: individual (RXJ1720.1+2638) -- 
galaxies: clusters: intracluster medium --
X-rays: galaxies: clusters -- radio continuum: general -- radio continuum: galaxies}

%

\section{Introduction}\label{sec:intro}

The high-resolution X-ray imaging capabilities of \chandra\ and {\em
  XMM-Newton} have provided an unprecedent view of galaxy clusters,
revealing a wealth of substructure in their cores and surrounding Mpc-scale
gaseous atmospheres.  In particular, {\em Chandra} showed that the
low-entropy gas in many, if not most, relaxed cool-core clusters is
``sloshing'' in the central potential well, generating the ubiquitous sharp,
arc-like gas density discontinuities, or ``cold fronts,'' that are
concentric with the cluster center and often form a spiral pattern 
\citep[e.g.,][see, for instance, Ghizzardi et al.\ 2010 for examples of cold fronts 
detected with {\em XMM-Newton}]{2001ApJ...562L.153M,2001ApJ...555..205M,2003ApJ...596..190M,
2003ApJ...583L..13D,2007PhR...443....1M, 2009ApJ...704.1349O, 2013A&A...555A..93E}.
Such sloshing motions are believed to result from a
recent gravitational perturbation of the cluster central potential in
response to collisions with small subclusters, which do not cause
significant visible X-ray disturbance outside the 
core \citep[e.g.,][]{2005ApJ...618..227T,2006ApJ...650..102A,2011ApJ...743...16Z, 2011MNRAS.413.2057R}. 
Active galactic nucleus (AGN) explosions in the cluster
central galaxy, occurring in an asymmetric gas distribution, may also
create a disturbance and set off sloshing of the core gas \citep{2001ApJ...562L.153M,2011MNRAS.415.3520H}.

A number of relaxed, cool-core clusters are hosts to radio ``minihalos'',
diffuse steep-spectrum\footnote{spectral index $\alpha >1$, for $S_{\nu}
  \propto \nu^{-\alpha}$, where $S_{\nu}$ is the flux density at the
  frequency $\nu$.} and low surface brightness radio sources, which enclose
-- albeit they are not obviously connected to -- the radio source associated
with the central elliptical galaxy \citep[e.g.,][and references therein]{2014ApJ...781....9G}.
Their emission typically fills the cooling region
($r\sim 50-300$ kpc) and often appears to be bounded by sloshing cold
fronts, suggesting a casual connection between minihalos and gas sloshing
\citep[S. Giacintucci et al. in preparation, M. Markevitch et al. in preparation]{2008ApJ...675L...9M,2013ApJ...777..163H,2014ApJ...781....9G}.
The origin of minihalos in cool-core clusters and their possible connection with
the giant radio halos found in merging clusters is still unclear \citep[e.g.,][for a review]{2014IJMPD..2330007B}. 
One possibility is that sloshing may
amplify the magnetic fields and induce turbulence in the cluster cool cores
\citep[hereafter Z13]{2004ApJ...612L...9F,2010ApJ...719L..74K,2011ApJ...743...16Z, 2012A&A...544A.103V,2013ApJ...762...78Z}.
Numerical simulations show that such turbulence
is generated mainly in the region enclosed by the cold fronts, with
velocities up to $\sim 200$ km s$^{-1}$ on scales of tens of kpc, whereas
negligible turbulence is driven outside the sloshing region (Z13).
Turbulence in the cool core, in turn, may re-accelerate pre-existing, aged
relativistic electrons in the intracluster medium (ICM) and, coupled with
the amplification of the local magnetic field, generate diffuse radio
emission within the cold front envelope with properties similar to the
observed minihalos \citep[turbulent reacceleration models,][Z13]{2002A&A...386..456G}.

As an alternative to turbulent reacceleration models, hadronic (or
secondary) models posit that the radio-emitting electrons in minihalos are
continuously injected by interactions between relativistic cosmic ray protons
with the cluster thermal proton 
population \citep{2004A&A...413...17P,2007ApJ...663L..61F,2010ApJ...722..737K,2010arXiv1011.0729K,2012ApJ...746...53F,2013MNRAS.428..599F,2014MNRAS.438..124Z}.
Recent numerical simulations of gas sloshing, modeling the formation of a
minihalo from secondary electrons emitting in the sloshing-amplified
magnetic field, have shown that, in these models, the radio emission 
is expected to be less confined within the sloshing region, due to the amplification
of the magnetic field in regions outside the cold fronts (ZuHone et al. 2014).
On average, the radio emission is found to be more extended than in
the turbulent reacceleration simulations, where the turbulence, and thus the
minihalo, are entirely confined to the region bounded by the cold fronts
(Z13).
\\
\\
In this paper, we present a radio/X-ray analysis of the cool-core cluster
RX\,J1720.1+2638 (hereafter RX\,J1720.1) at $z=0.16$, which is host 
to a radio minihalo in its
center.  This cluster was the first relaxed system in which sloshing cold
fronts have been revealed by {\em Chandra} \citep{2001ApJ...555..205M} as well
as one of the first two clusters in which a correlation between minihalo and
cold fronts has been reported \citep{2008ApJ...675L...9M}. Here, we use
multi-frequency radio observations from the Giant Metrewave Radio Telescope
(\gmrt) and Very Large Array (\vla) to study the spectral properties of the
minihalo, which provide important information on the origin of the
radio-emitting electrons, and investigate its connection with the sloshing
cold fronts seen in the {\em Chandra} image.

In Table 1, we summarize the general properties of RX\,J1720.1. We 
adopt $\Lambda$CDM cosmology with H$_0$=70 km s$^{-1}$ Mpc$^{-1}$, 
$\Omega_m=0.3$ and $\Omega_{\Lambda}=0.7$.


\begin{table}[t]
\caption[]{Properties of the galaxy cluster RX\,J1720.1+2638}
\begin{center}
\begin{tabular}{lc}
\hline\noalign{\smallskip}
\hline\noalign{\smallskip}
Parameter & Value \\
\hline\noalign{\smallskip}
$^{\rm a}$ R.A.$_{\rm J2000}$ (h m s)  &  17 20 09.3 \\ 
$^{\rm a}$ Decl.$_{\rm J2000}$ ($^{\circ}$ $^{\prime}$ $^{\prime\prime}$) & +26 37 38 \\
\phantom{0}$z$ &  0.16   \\
\phantom{0}$D_L$ (Mpc) & 765.4 \\
\phantom{0}Linear scale (kpc/$^{\prime \prime}$) &  2.758    \\ 
$^{\rm b}$ $L_{\rm X,\,500 \, [0.1-2.4 \, keV]}$ ($10^{44}$ erg s$^{-1}$) & $7.1$ \\
$^{\rm c}$ $kT$ (keV) & $6.3$ \\
$^{\rm d}$ $M_{\rm 500}$ ($10^{14}$ $M_{\odot}$) & $6.3$  \\
\noalign{\smallskip}
\hline\noalign{\smallskip}
\end{tabular}
\end{center}
\label{tab:sources}
{\bf Notes.} 

$^{\rm a}$ J2000 X-ray coordinates from \cite{2011A&A...534A.109P}.

$^{\rm b}$ $[0.1-2.4]$ keV X-ray luminosity within $R_{\rm 500}$ from 
\cite{2011A&A...534A.109P}, where $R_{\rm 500}$ is the radius corresponding 
to a total density contrast $500\rho_c(z)$, $\rho_c(z)$ 
being the critical density of the Universe at the cluster redshift.

$^{\rm c}$ Global cluster temperature from \cite{2009ApJS..182...12C}. 

$^{\rm d}$ Cluster mass within $R_{500}$ from Planck collaboration et al. (2013).
\end{table}



\begin{table*}
\caption[]{Radio observations of RX\,J1720.1+2638}
\begin{center}
\begin{tabular}{lccccccccc}
\hline\noalign{\smallskip}
\hline\noalign{\smallskip}
Array & Project & Frequency & Bandwidth & Observation &
Time  & FWHM, p.a.  &   rms  &  $u-v$ range & $\theta_{\rm LAS}$  \\
          &         &  (GHz)    &   (MHz)   &   date   &  
(min) & ($^{\prime \prime} \times^{\prime \prime}$, $^{\circ}$)\phantom{00} & ($\mu$Jy b$^{-1}$) & (k$\lambda$) & ($^{\prime}$) \\
\noalign{\smallskip}
\hline\noalign{\smallskip}
{\em GMRT} &  11MOA01 & 0.317 & \phantom{0}12$^{\rm a}$ & 2007 Mar 8 & 220 & $9.1\times8.0$, 4 & 210 & 0.15-25.2 & 14 \\
{\em GMRT} & 11MOA01 & 0.617 & \phantom{0}11$^{\rm b}$ &  2007 Mar 10 & 250 & $5.0\times4.3$, $-71$ & 28 & 0.22-53.5 & 9\\
{\em GMRT} & 11MOA01 & 1.28 & 21 & 2007 Mar 8 & 320 & $2.5\times2.2$, 87 & 45 & 0.41-112 & 5 \\
{\em VLA}--BnA & AH988$^{\rm c}$ & 1.42 & 25 & 2009 Jan 26 & 60 & $3.8\times2.1$, 87 & 40 & 1.1-80 & 2\\ 
{\em VLA}--A & AE117 & 1.42 & 50 & 1998 Apr 12 & 20 &  $1.5\times1.3$, 68 & 20 & 3.3-166 & 0.6 \\
{\em VLA}--B & AH190 & 1.48 & 25 &  1985 Apr 25 & 70 & $4.6\times3.7$, 68 & 30 & 0.5-52.5 & 4 \\
{\em VLA}--A & AF233 & 4.86 & 50 & 1992 Oct 20 & 1 & $0.8\times0.4$, 66 & 70 & 16-550 & $\sim 0.05^{\rm d}$\\
{\em VLA}--B & AH190 & 4.86 & 50 & 1985 Apr 25 & 30 & $1.4\times1.2$, 64 & 30 & 2.1-181 & 1 \\
{\em VLA}-C & AE125 & 4.86 & 50 & 1999 Jan 16 & 4 & $4.1\times3.6$, $-13$ & 40 & 0.7-53 & $\sim 2.5^{\rm d}$\\
{\em VLA}-DnC & AH0355 & 8.44 & 50 & 1989 Jun 2 &  3 & $6.1\times2.6$, 78 & 35 & 1.2-62 & $\sim1.5^{\rm d}$\\
\hline{\smallskip}
\end{tabular}
\end{center}
\label{tab:obs}
{\bf Notes.} Column 1: radio telescope. Column 2: project code. Columns 3--5: frequency, usable bandwidth after
bandpass calibration, and observation date. Column 6: useful time on source after flagging. Columns 7 and 8: full 
width at half-maximum (FWHM) and position angle (PA) of the synthesized beam and rms noise level ($1\sigma$) in 
images made using a uniform weighting scheme ($ROBUST=-5$). Columns 9: effective $u-v$ range of the observation.
Column 10: largest angular scale detectable by the array.
\\
$^{\rm a}$ The observation was made using both USB (central frequency 333 MHz) and LSB (central frequency 317 MHz) 
with an observing bandwidth of 16 MHz each (before bandpass calibration), but only the LSB dataset was used 
for the analysis presented in this paper (see \S \ref{sec:gmrt} for details).
\\
$^{\rm b}$ The observation was made using both USB (central frequency 617 MHz) and LSB (central frequency 602 MHz) 
with an observing bandwidth of 16 MHz each (before bandpass calibration), but only the USB dataset was used for 
the analysis presented in this paper (see \S \ref{sec:gmrt} for details).
\\
$^{\rm c}$ An image from this observation has been presented by \cite{2012AJ....144...48H}.
\\
$^{\rm d}$ Due to the short duration of this observation, the angular scale that can be imaged reasonably 
is much smaller than the nominal $\theta_{\rm LAS}$ of full-synthesis observations in the 
same array configuration 
($0.15^{\prime}$ for {\em VLA}--A at 4.9 GHz, $5^{\prime}$ for {\em VLA}--C at 4.9 GHz and $3^{\prime}$ for 
{\em VLA}--DnC at 8.4 GHz; http://science.nrao.edu/facilities/vla/proposing/oss/ossjan09.pdf). 
\end{table*}



\begin{table*}
\caption[]{Properties of the Radio Galaxies}
\begin{center}
\begin{tabular}{lccccccccc}
\hline\noalign{\smallskip}
\hline\noalign{\smallskip}
Radio Source & $S_{\rm 317 \, MHz}$ & $S_{\rm 617 \, MHz}$ & $S_{\rm 1.28 \, GHz}$ & $S_{\rm 1.48 \, GHz}$ &  $S_{\rm 4.86 \, GHz}$ &  $S_{\rm 8.44 \, GHz}$ & $\alpha_{\rm tot}$ & $P_{\rm 1.48 \, GHz}$ & Size \\
  & (mJy) & (mJy) & (mJy) & (mJy) & (mJy) & (mJy) &  & ($10^{24}$ W Hz$^{-1}$) & (kpc) \\
\noalign{\smallskip}
\hline\noalign{\smallskip}
point source (BCG) & \phantom{0}$24\pm2^{\rm a}$ & \phantom{0}$11\pm1^{\rm a}$ & $6.9\pm0.4$ & $6.7\pm0.3$ & \phantom{0}$2.3\pm0.1^{\rm b}$ & $1.4\pm0.1$ & $0.87\pm0.03$ & $0.47\pm0.02$  & $<1.4^{\rm c}$\\
head tail & $31\pm3$ & $12\pm1$ & $5.4\pm0.3$ & $5.5\pm0.3$ & $1.9\pm0.1$ & $1.0\pm0.1$ & $1.05\pm0.04$ & $0.39\pm0.02$ & \phantom{0}$140^{\rm d}$  \\
wide-angle tail & $72\pm6$ & $44\pm2$ & $30\pm2$ & $27\pm1$ & $-$ & $-$ & $0.64\pm0.06$  & $1.89\pm0.09$ & \phantom{0}$500^{\rm d}$ \\   \hline{\smallskip}
\end{tabular}
\end{center}
\label{tab:flux}
{\bf Notes.} Column 1: radio source. Columns 2--7: radio flux densities measured at full resolution (uniform weighting; Table \ref{tab:obs}). Column 8: total spectral index. Column 9: radio power at 1.48 GHz. Column 10: largest linear size.
\\
$^{\rm a}$ Measured on images obtained using only baselines $>15$ k$\lambda$.
\\
$^{\rm b}$ From the {\em VLA} B-configuration image (Fig.~\ref{fig:mh_center}b).
\\
$^{\rm c}$ Beam-deconvolved size from a Gaussian fit to the source in the {\em VLA} 
A-configuration image at 4.9 GHz (Fig.~\ref{fig:mh_center}a).
\\
$^{\rm d}$ Measured on the 617 MHz image (Fig.~\ref{fig:field}).
\end{table*}

%
%
\begin{figure*}
\centering
\includegraphics[width=0.89\textwidth,bb=40 70 450 397, clip]{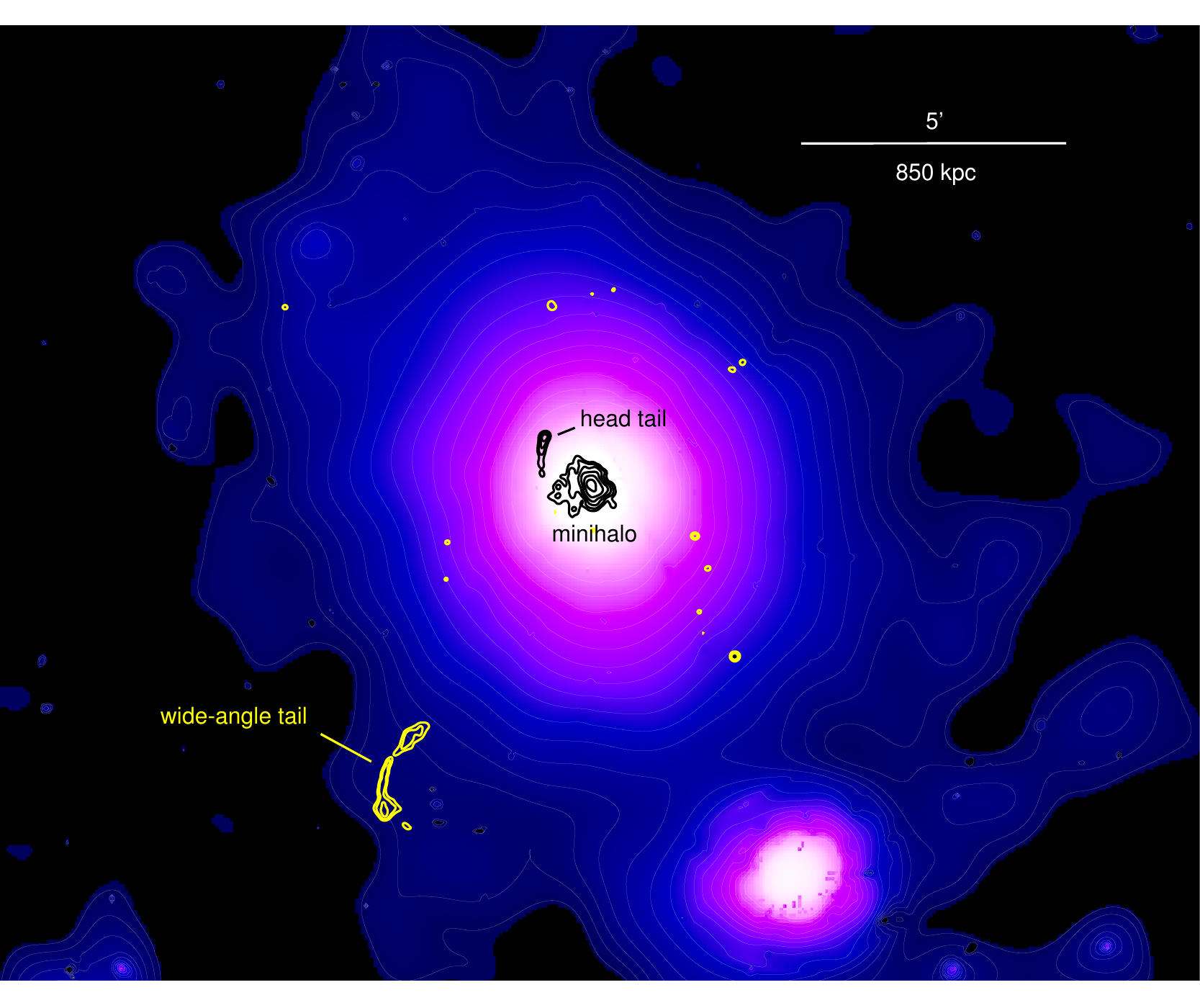}
\smallskip
\caption{Radio and X-ray emissions in RX\,J1720.1. The size of the field 
is $18^{\prime}\times15^{\prime}$ (3 Mpc $\times$ 2.5 Mpc). The GMRT
617 MHz image at a resolution of $5.6^{\prime\prime}\times4.7^{\prime\prime}$, in p.a. $-74^{\circ}$
is shown as black and yellow contours, spaced by a factor of 2 starting from 0.2 mJy beam$^{-1}$.
The extended radio sources are labelled. The X-ray image (color and white contours)
is a wavelet reconstruction of the \xmm\ point source-subtracted 
image in the 0.5-2.5 keV band. The X-ray contours are spaced by a factor of $\sqrt2$.}
\label{fig:field}
\end{figure*}
%
%

%
%
\begin{figure*}
\centering
\hspace{-0.5cm}
\includegraphics[width=17cm]{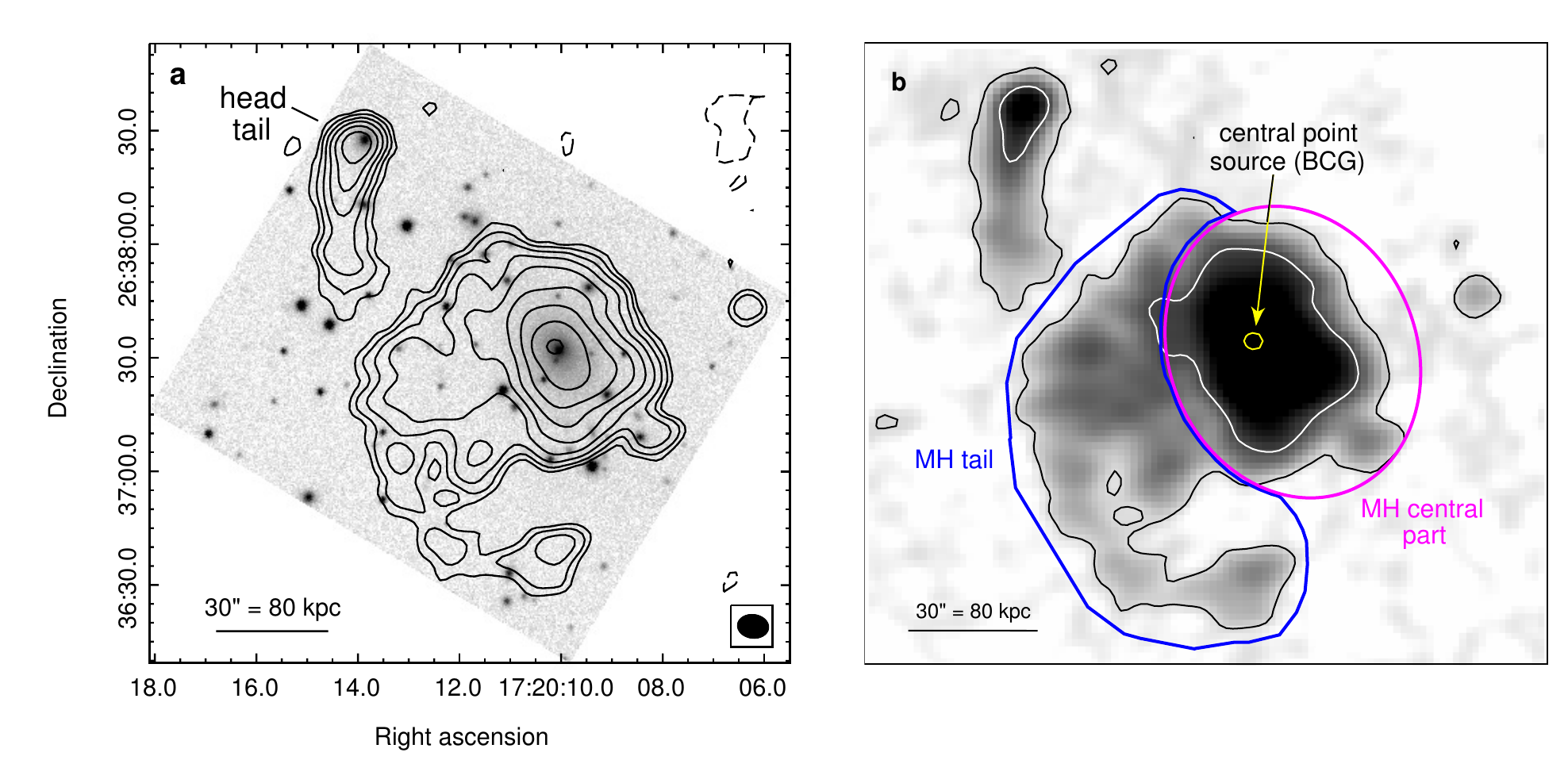}
\smallskip
\caption{(a) \gmrt\ 617 MHz contours of the central minihalo in RX\,J1720.1 
and nearby head-tail radio source, associated with a cluster member galaxy 
at $\sim 1^{\prime}.3$ from the BCG (see \S~\ref{sec:ps}). The radio image 
has been obtained using natural weighting and is overlaid on the optical r-band 
SDSS image. The restoring beam (black ellipse) is $7^{\prime\prime}.8\times6^{\prime\prime}.1$, 
in p.a. $-83^{\circ}$ and r.m.s. noise level is $1\sigma=30$ $\mu$Jy beam$^{-1}$. 
Contour levels are spaced by a factor of 2 starting from $+3\sigma$. Contours
at $-3\sigma$ are shown as dashed. (b) Grayscale image at 617 MHz (same as (a))
with contours at 0.09 (black) and 1.4 (white) mJy beam$^{-1}$. The minihalo is composed 
by a bright central part (magenta region) and a much fainter tail to the south-east 
(blue region; \S~\ref{sec:mh}). The yellow circle marks the position of the
radio point source associated with the BCG.}
\label{fig:scheme}
\end{figure*}
%
%

%
%
\begin{figure*}
\centering
\includegraphics[width=17cm]{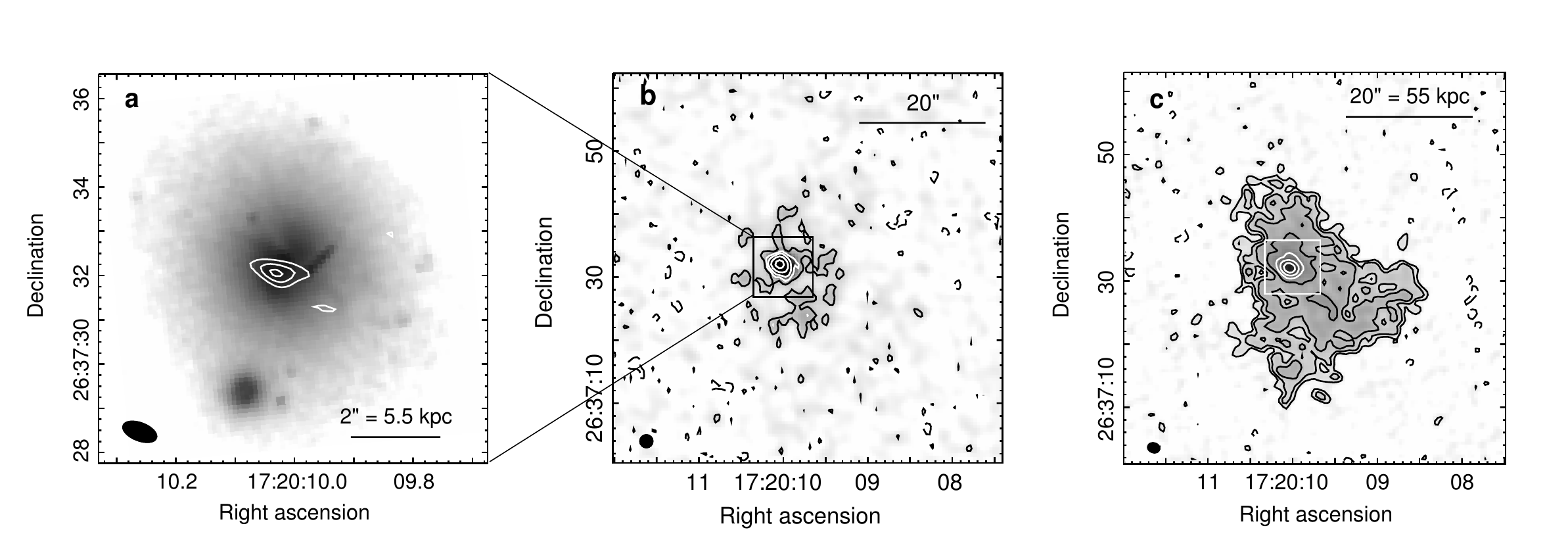}
\smallskip
\caption{(a) \vla\ A-configuration contours at 4.86 GHz, overlaid on the {\em HST} WFPC2 
image of the BCG (grayscale). The restoring beam (black ellipse) is $0^{\prime\prime}.8\times0^{\prime\prime}.4$ , 
in p.a. $66^{\circ}$ and r.m.s. noise level is $1\sigma=70$ $\mu$Jy beam$^{-1}$. 
Contours are 0.2, 0.4, 0.6 mJy beam$^{-1}$. (b) \vla\ B-configuration image at 4.86 GHz (grayscale and contours)
and (c) \vla\ A-configuration image at 1.42 GHz (grayscale and contours) of the point source at the BCG 
(white contours) and innermost region of the diffuse minihalo (black contours). 
The restoring beam is $2^{\prime\prime}$ (black circle) and $1^{\prime\prime}.9\times1^{\prime\prime}.5$, 
in p.a. $63^{\circ}$ (black ellipse), respectiveley. The r.m.s. noise levels are 
$1\sigma=30$ $\mu$Jy beam$^{-1}$ and $1\sigma=15$ $\mu$Jy beam$^{-1}$. 
Contours are spaced by a factor of 2 from $+3\sigma$. Contours at $-3\sigma$ are 
shown as dashed. The central box indicates the region covered by the image in panel (a).}
\label{fig:mh_center}
\end{figure*}
%
%

\section{Radio observations}\label{sec:obs}

We obtained radio observations of RX\,J1720.1 
from the \gmrt\ and \vla\ archives, covering an 
interval of almost two orders of magnitude in frequency 
($317 \, {\rm MHz}\div8.4$ GHz) with data at 
6 different frequencies. We summarize these observations in 
Table 2.

\subsection{{\em GMRT} observations}\label{sec:gmrt}

RX\,J1720.1 was observed with the \gmrt\ for a total of 
about 6 hours at 325 MHz, 7 hours at 610 MHz and 7 hours 
at 1.28 GHz, including calibration overheads (project 11MOA01; 
Table 2). The observations were made in 
spectral-line observing mode, using the \gmrt\ hardware backend.
The upper and lower side bands (USB and LSB) were used simultaneously, 
for a total observing bandwidth of 32 MHz, divided in 256 channels. 

We calibrated and reduced these observations using the NRAO\footnote{National Radio Astronomy Observatory.} 
Astronomical Image Processing System (AIPS) package,
as described in \cite{2008A&A...486..347G,2011ApJ...732...95G}.
We found that all data sets were in part impacted by 
radio frequency interference (RFI). Therefore, we accurately inspected and 
edited the data to remove the RFI-affected visibilities, 
leaving approximately 5 hours of usable time at 325 MHz and
4 hours at 610 MHz and 1.28 GHz each. After the initial amplitude  
calibration and bandpass calibration obtained using the primary flux density calibrators,  
we averaged the central channels in the USB and LSB data 
sets at 325 MHz to 12 channels/band, with each channel $\sim 1$ MHz wide. 
At 610 MHz and 1.28 GHz, we averaged each band to 6 channels of 1.75 MHz 
width each. We then used the phase calibrators to calibrate the data in phase 
and finally applied a number of phase-only self-calibration cycles and 
imaging to the target visibilities to correct residual phase errors. 
We used wide-field imaging in each step of the self-calibration 
process. The images from the USB and LSB data sets were combined to obtain
the final images at each frequency, which were then corrected
for the \gmrt\ primary beam response\footnote{see http://www.ncra.tifr.res.in:8081/{\textasciitilde}ngk/primarybeam/beam.html for
details on the GMRT primary beam shapes.} using PBCOR in AIPS. 

Due to residual phase instabilities in the USB data at 325 MHz and 
LSB data at 610 MHz, the band combination led to images with a quality 
worse than those obtained from the LSB and USB data sets alone, respectively. 
We thus used only the LSB data, centered at 317 MHz, and USB data, centered at 617 MHz, 
for the analysis presented in this paper. Table 2 summarizes the restoring 
beams and rms noise levels ($1\sigma$) of the final images obtained 
with the Briggs ``robustness'' parameter set to $ROBUST=-5$ in IMAGR
(uniform weighting).

At all frequencies, the sources 3C\,147, 3C\,286 and 3C\,48 were used as 
flux density calibrators. Their flux density was set using the VLA Perley \& Taylor 
(1999.2) values, adopted as a default in the release 31DEC10 of 
the task SETJY. Residual amplitude errors are estimated to be within $5\%$ at 610 
MHz and 1.28 GHz and $8\%$ at 317 MHz \citep{2004ApJ...612..974C}.

\subsection{{\em VLA} observations}

RX\,J1720.1 was observed with the {\em VLA}\footnote{Old, pre-WIDAR correlator VLA.} at
1.4 GHz using the A and BnA configurations for 
20 minutes and 1 hour, respectively, 
and at 1.5 GHz (70 minutes) and 4.86 GHz (30 minutes) 
in B configuration (Table 2). Much shorter observations (few minutes) were made 
at 4.86 GHz in A and C configurations and 8.44 GHz in DnC (Table 2).

We calibrated and reduced all data sets in AIPS following standard procedures
and applied self-calibration in phase to reduce the effects of residual phase
variations in the data and improve the quality of the final images. 
These were then corrected for the primary beam attenuation of the {\em VLA} antenna 
using the task PBCOR. The rms noise level ($1\sigma$) achieved in 
the final images made using the uniform weighting scheme are summarized in Table 2.

At all frequencies, the flux density scale was set using 3C\,286 (and
3C\,147 in project AH988) and the VLA Perley \& Taylor (1999.2) values in SETJY.
A conservative estimate of the amplitude calibration errors, based on the 
residual antenna gains and on the flux density of the primary calibrator, is
of the order of $\sim 4-5$ \% at all frequencies.

\section{The field of RX\,J1720.1}\label{sec:field}

Figure \ref{fig:field} presents a $15^{\prime}\times18^{\prime}$ (3 Mpc $\times$ 2.5 Mpc) region
containing RX\,J1720.1. Black and yellow contours are the \gmrt\ 617 MHz image at a resolution of 
$\sim 5^{\prime\prime}$ FWHM and the color image is a wavelet reconstruction of the 
{\em XMM-Newton} X-ray image in the 0.5-2.5 keV band (from ObsIDs 0500670201, 0500670301 and 0500670401).
On this large scale, the cluster has a relaxed and regular X-ray morphology with a bright central core.

In the radio, the field is dominated by three extended radio sources -- the diffuse minihalo 
in the cluster core (\S\ref{sec:mh}), a nearby head-tail source at a projected distance of $1^{\prime}.3$ 
from the center, and a wide-angle tail at the cluster outskirts. Both tailed sources are associated 
with cluster member galaxies with coordinates R.A.$_{\rm J2000}$= 17h 20m 13.9s and 
Decl.$_{\rm J2000}=+26^{\circ} \, 38^{\prime} \, 28^{\prime\prime}$ 
and R.A.$_{\rm J2000}$= 17h 20m 27.5s and Decl.$_{\rm J2000}=+26^{\circ} \, 31^{\prime} \, 59^{\prime\prime}$
and redshifts $z=0.163$ and $z=0.159$, respectively
(Owers et al. 2011). Their radio properties are summarized in Table 2.
Radio luminosity, spectral index and size are within the range of values commonly reported 
for this type of cluster radio galaxies \citep[e.g.,][]{2002ASSL..272..163F}.

It is worth noticing that wide-angle tails are typically associated with cluster/group central 
galaxies. This suggests that the wide-angle tail at the periphery of RX\,J1720.1 may 
reside at the center of an infalling subcluster. Inspection of the galaxy distribution in 
\cite{2011ApJ...741..122O} reveals a weak optical peak at the location of the wide-angle tail.


\begin{table*}
\caption[]{Properties of the Minihalo}
\begin{center}
\begin{tabular}{cccccccccccc}
\hline\noalign{\smallskip}
\hline\noalign{\smallskip}
Minihalo & $S_{\rm 317 \, MHz}$ & $S_{\rm 617 \, MHz}$ & $S_{\rm 1.28 \, GHz}$ & $S_{\rm 1.48 \, GHz}$ &  $S_{\rm 4.86 \, GHz}$ &  $S_{\rm 8.44 \, GHz}$ & $\alpha_1$ & $\alpha_2$ & $P_{\rm 1.48 \, GHz}$ & Size \\
 region     & (mJy) & (mJy) & (mJy) & (mJy) & (mJy) & (mJy) &  & &($10^{24}$ W Hz$^{-1}$) & (kpc) \\
\noalign{\smallskip}
\hline\noalign{\smallskip}

total  &  $365\pm58$ & $170\pm12$ & $65\pm4$ & $68\pm5$ & $20.3\pm1.5$ & $6.6\pm0.7$ &  $1.1\pm0.1$ & $2.0\pm0.3$ & $4.8\pm0.4$ & \phantom{0}$r\sim140$ \\

center &  $286\pm38$ & $144\pm11$ & $59\pm3$ & $60\pm5$ & $18.7\pm1.3$ & $6.2\pm0.6$ & $1.0\pm0.1$ & $2.0\pm0.2$ & $4.2\pm0.4$ & $r\sim80$ \\

tail &  $79\pm6$ & $26\pm2$ & \phantom{0}$6\pm1$ & $8\pm1$ & \phantom{0}$1.6\pm0.5$ & $>0.4$  & $1.4\pm0.1$ & $<2.5$ &  $0.56\pm0.07$  & \phantom{00}$l\sim230^{\rm a}$ \\
\hline{\smallskip}
\end{tabular}
\end{center}
\label{tab:flux_mh}
{\bf Notes.} Column 1: region of the minihalo. Column 2--7: radio flux densities measured from Figs.~\ref{fig:scheme} and \ref{fig:mh_full}, using the magenta and 
blue regions shown in Fig.~\ref{fig:scheme}. Column 8: spectral index between 317 MHz and 4.86 GHz. Column 9: spectral index between 4.86 GHz and 8.44 GHz. 
Column 10: radio power at 1.48 GHz. Column 11: linear size ($r$: radius, $l$: length).  
\\
$^{\rm a}$  Measured on the 617 MHz image (Fig.~\ref{fig:scheme}). 

\end{table*}

\section{Radio emission in the cluster core}\label{sec:core}

In Fig.~\ref{fig:scheme}, we zoom on the cluster center, 
showing a 617 MHz image at the resolution of $\sim 8^{\prime\prime}$, 
obtained using natural weighting ($ROBUST=5$) to enhance the extended emission. 
Panel (a) presents an overaly of the radio contours on the optical r-band 
SDSS\footnote{Sloan Digital Sky Survey.} image and panel (b) shows the same radio image as 
grayscale. The yellow circle indicates the position of a point source, 
which is coincident with the brightest cluster galaxy (BCG; see \S\ref{sec:ps}). 
The surrounding diffuse minihalo is composed by a bright central part (magenta region) 
and a lower surface brightness, arc-shaped ``tail'' to the south-east 
(blue region; \S\ref{sec:mh}). Also visible in the image 
is the head-tail radio galaxy.

\subsection{The central radio galaxy}\label{sec:ps}

The radio source associated with the BCG (R.A.$_{\rm J2000}$= 17h 20m 10.0s, 
Decl.$_{\rm J2000}=+26^{\circ} \, 37^{\prime} \, 32^{\prime\prime}$; \cite{2011ApJ...741..122O})
is unresolved in all existing observations of RX\,J1720.1. The VLA A-configuration 
data set at 4.86 GHz provides the highest angular resolution radio image ($0^{\prime\prime}.8\times0^{\prime\prime}.4$ FWHM), 
which is presented in Fig.~\ref{fig:mh_center}(a), overlaid on the optical 
{\em HST} WFPC2\footnote{{\em Hubble Space Telescope} Wide-Field Planetary Camera 2.} 
image. A single, compact component coincides with the optical peak of the galaxy; 
its beam-deconvolved size is $<0^{\prime\prime}.5$, which implies a linear 
size $<1.4$ kpc. 

In Figs.~\ref{fig:mh_center}(b) and (c), we present  \vla\, $2^{\prime\prime}$--resolution images 
at 4.86 GHz and 1.42 GHz of the central $160\times160$ kpc$^2$ region of the cluster. 
In these images, the point source at the BCG (white contours) is enshrouded by the innermost part of 
the larger-scale minihalo. No obvious extended features connected to the central point source, 
such as jets or lobes, are visible.

The BCG radio properties are summarized in Table 3. All flux densities were
measured on images obtained using the uniform weighting scheme (Table 2) 
by fitting the source with a Gaussian model
(task JMFIT in AIPS). At 317 MHz and 617 MHz, where it is not possible to separate well 
the point source from the minihalo due to the lower angular resolution, we used 
images obtained cutting the innermost $15$ k$\lambda$ region of the $u-v$ plane 
to suppress the larger-scale emission and image the point source alone.

\subsection{The radio minihalo}\label{sec:mh}

\begin{figure*}
\centering
\includegraphics[width=7cm]{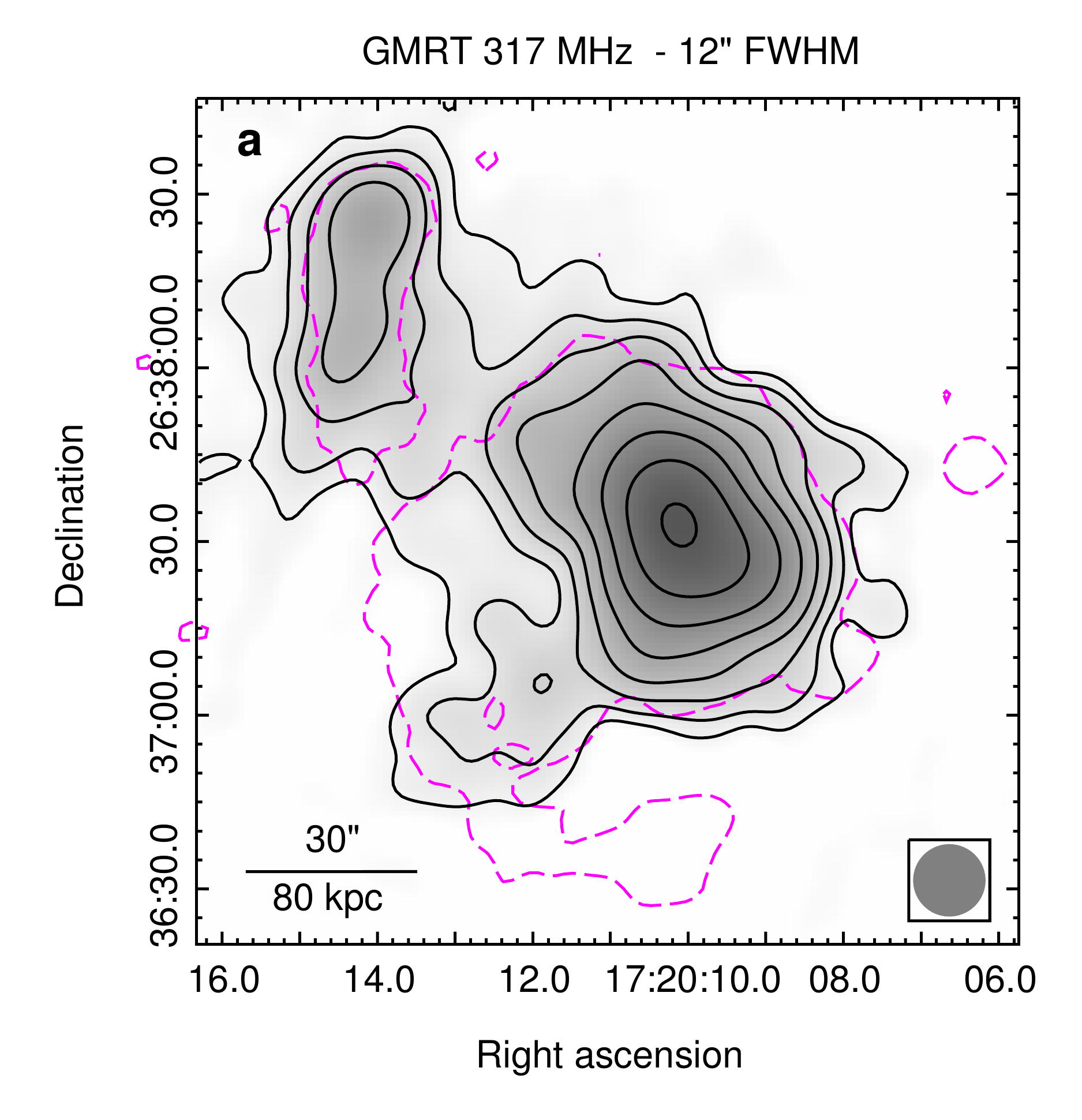}
\includegraphics[width=7cm]{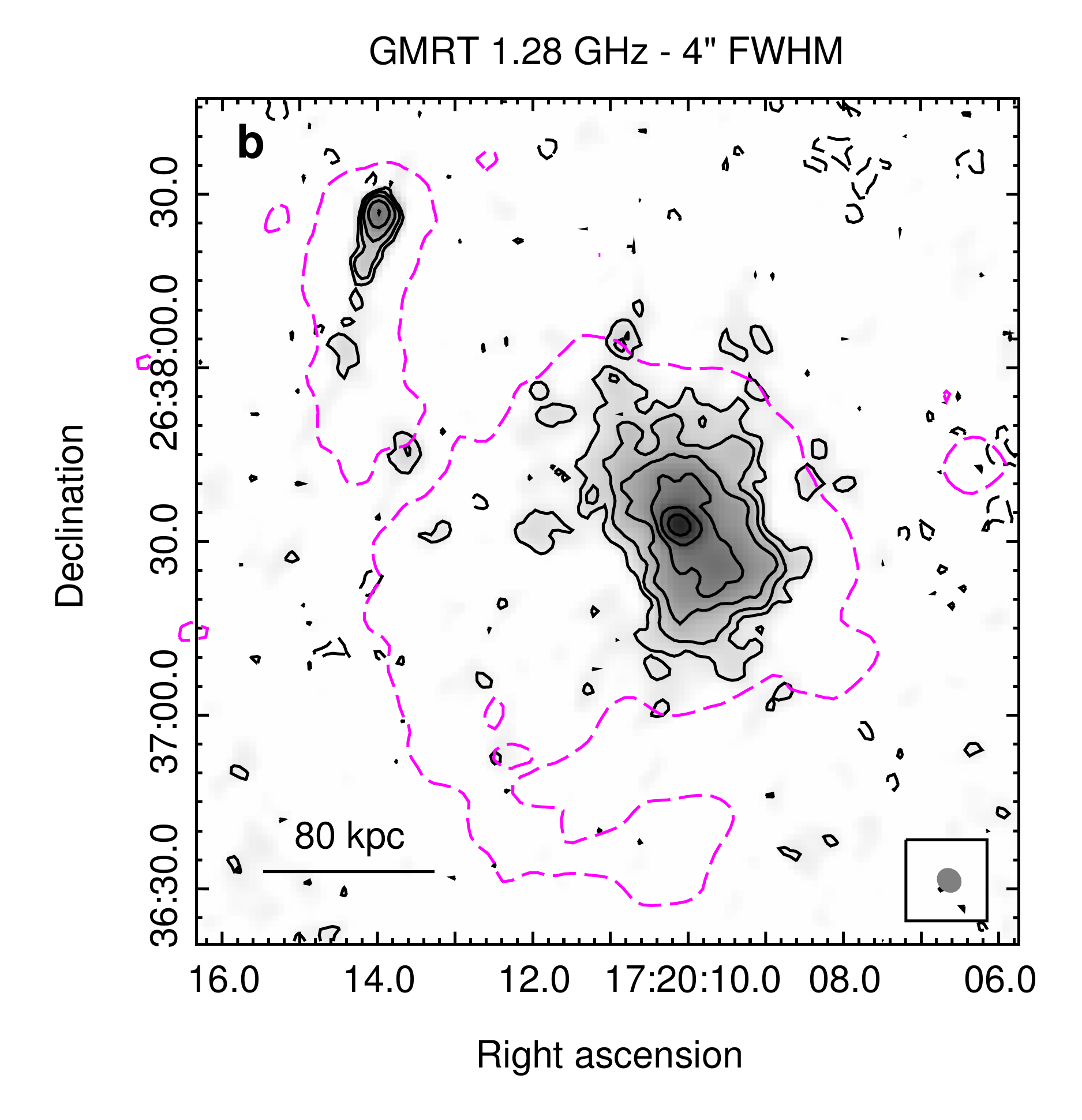}
\includegraphics[width=7cm]{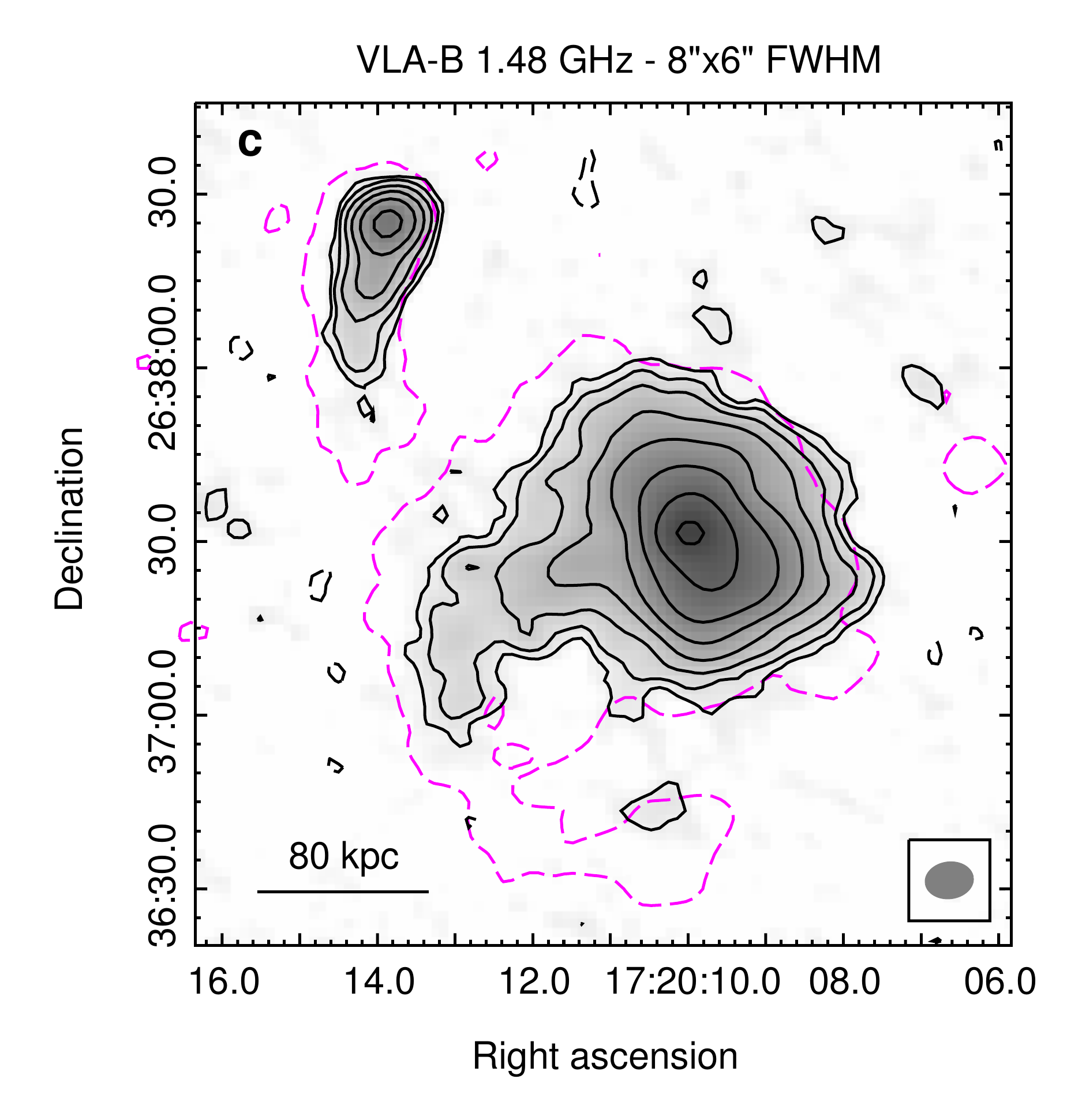}
\includegraphics[width=7cm]{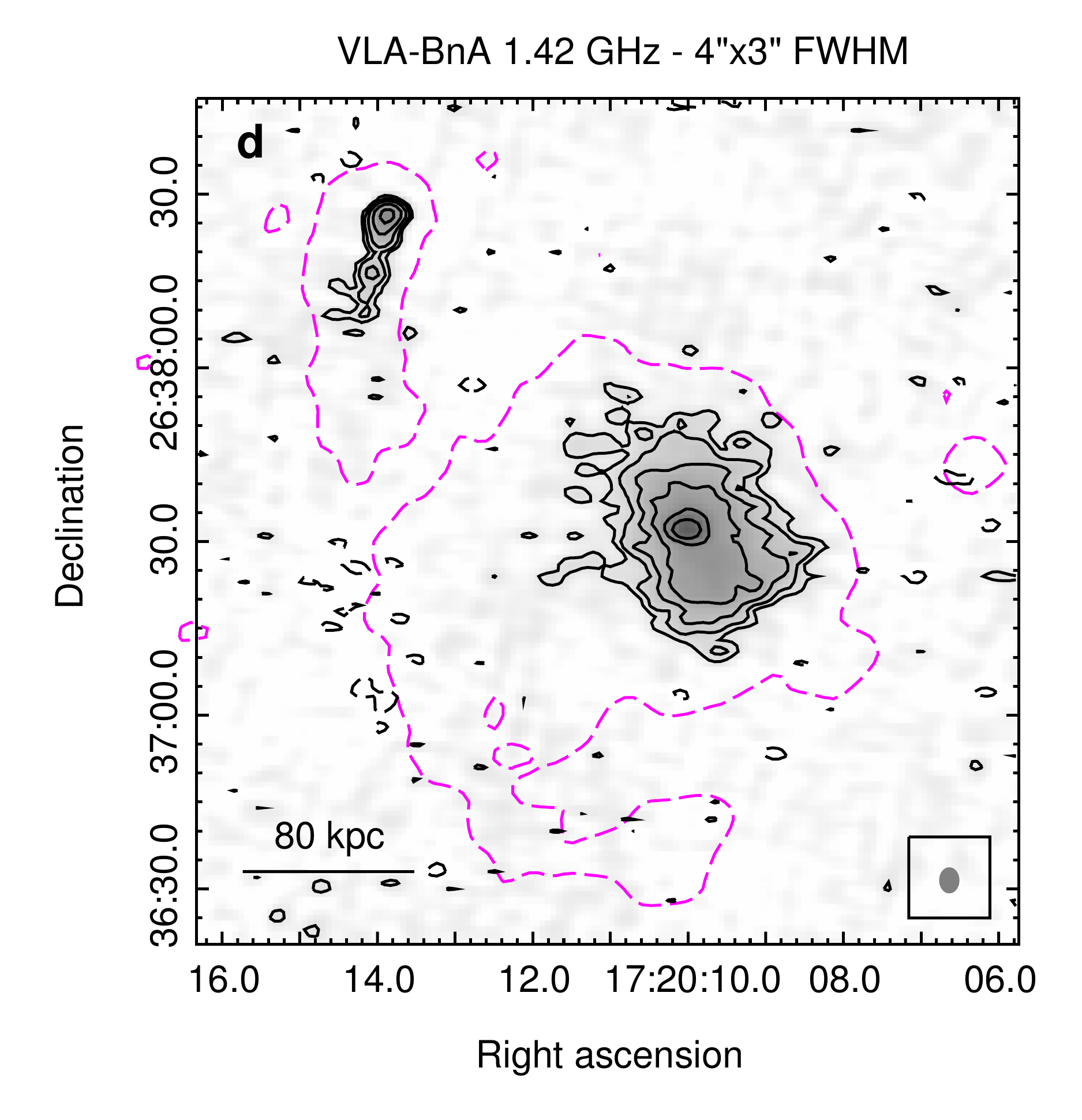}
\includegraphics[width=7cm]{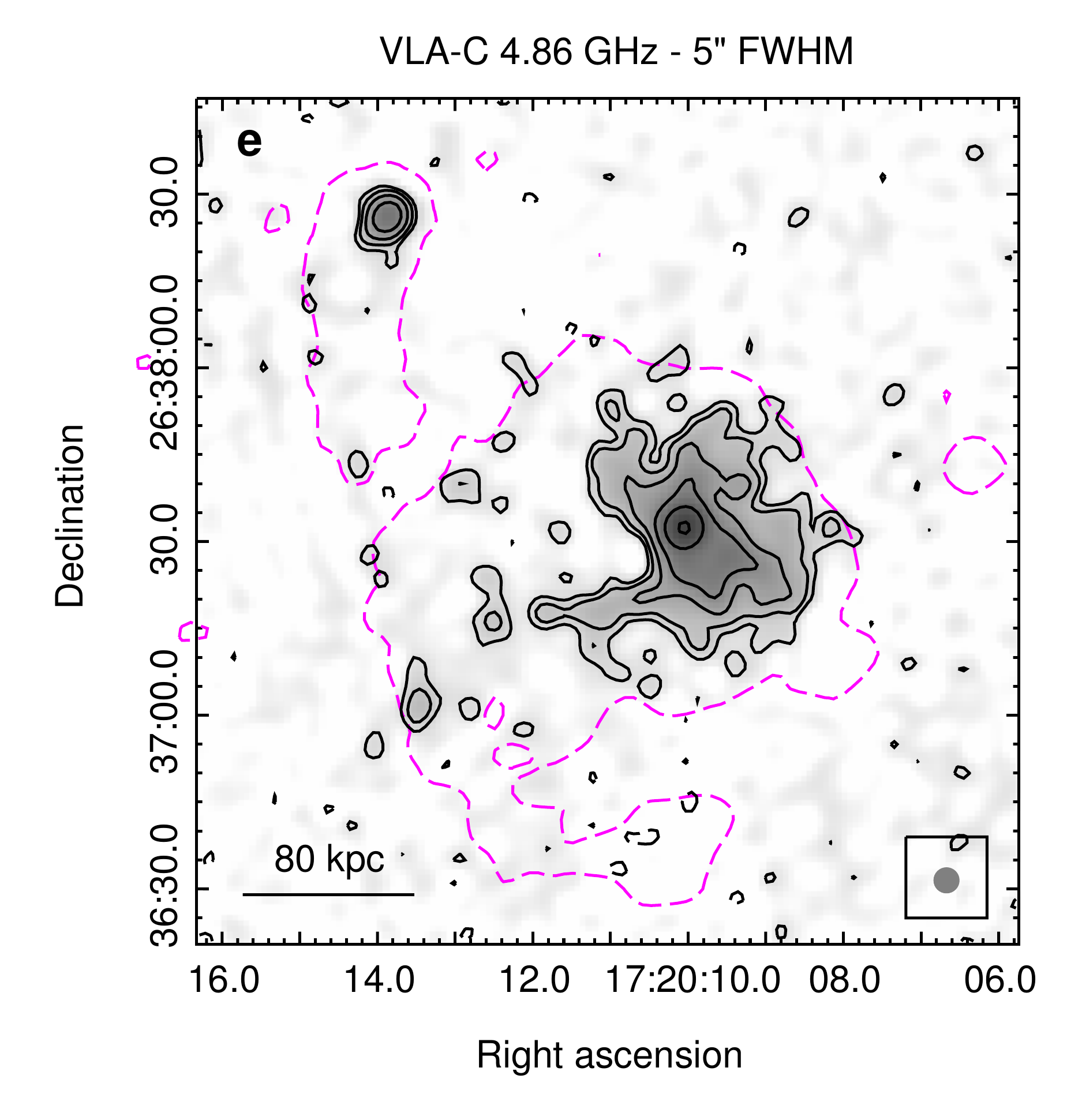}
\includegraphics[width=7cm]{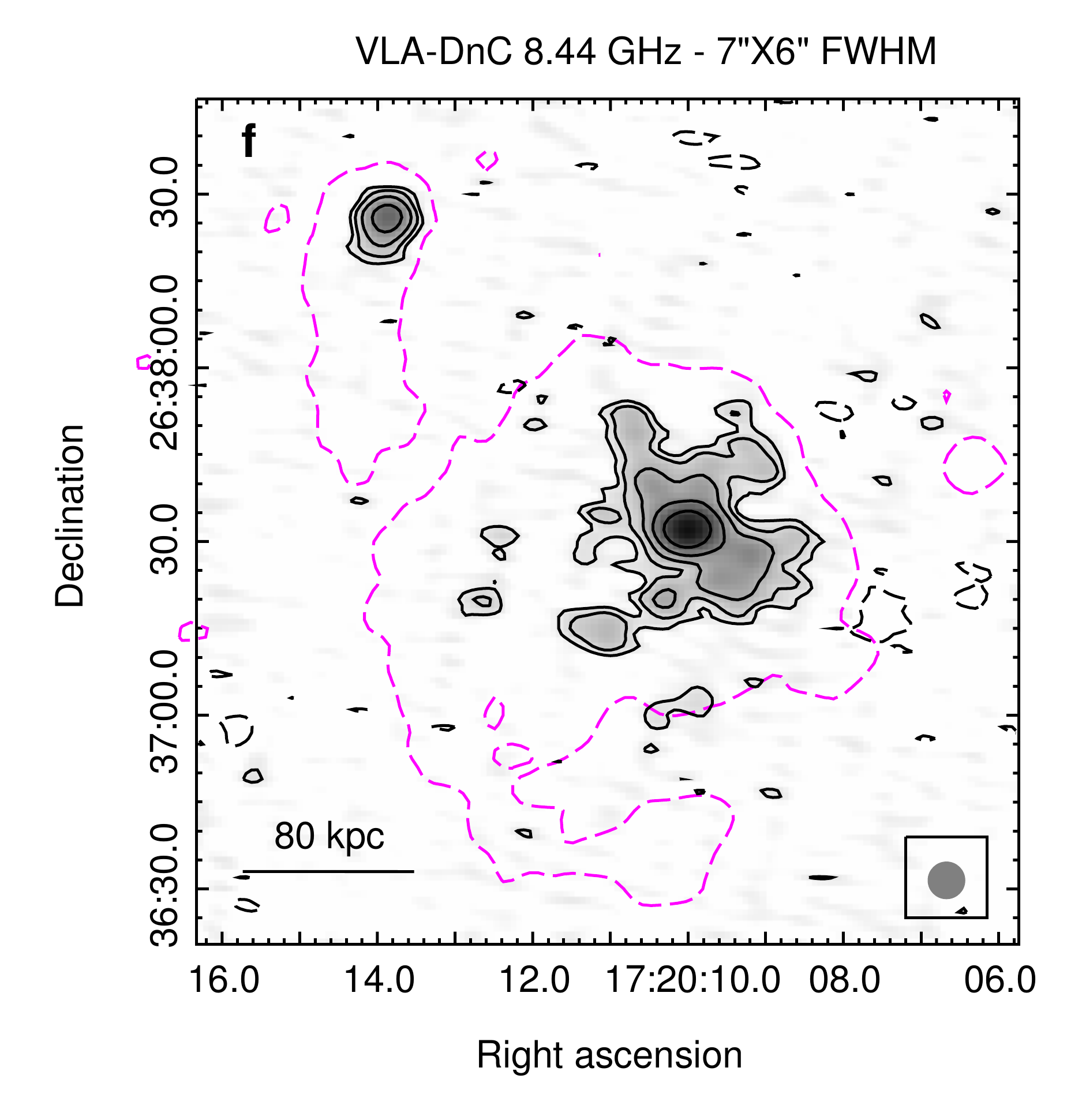}
\smallskip
\caption{Radio images (grayscale and black contours) of the minihalo at 317 MHz (a), 
1.28 GHz (b), 1.42 GHz (c), 1.48 GHz (d), 4.86 GHz (e) and 8.44 GHz (f),
obatined using natural weighting. The gray ellipse
in the lower right corner of each image shows the radio beam, whose FWHM
is reported at the top of the corresponding panel. The r.m.s. noise levels 
are 220, 50, 40, 30, 40 and 35 $\mu$Jy beam$^{-1}$, respectively. Black contours are spaced by a 
factor of 2 starting from $+3\sigma$. When present, contours 
corresponding to the $-3\sigma$ level are shown as black dashed. For
a comparison with the 617 MHz image in Fig.~\ref{fig:scheme}, the 
lowest contour at 617 MHz is reported in magenta in all panels.} 
\label{fig:mh_full}
\end{figure*}

At all frequencies, we made images of the minihalo over a range of resolutions, 
using different weighting schemes by varying the $ROBUST$ parameter in IMAGR, to thoroughly 
check the reliability of the features and the robustness of the flux density measurements. We
obtained images ranging from uniform weights ($ROBUST=-5$) to a scheme close to natural 
weighting ($ROBUST=5$). To better highlight the diffuse structure, in Fig.~\ref{fig:mh_full}, 
present the naturally-weighted images of the minihalo at increasing frequency from 317 MHz (a) to 8.44 GHz (f). 
The angular resolution of these images  -- ranging from $4^{\prime\prime}$ at 1.28 GHz to $12^{\prime\prime}$ at 317 MHz -- 
is slightly lower than that of the corresponding uniformly-weighted images and r.m.s. noise 
levels are similar (see figure caption and Table 2). For a comparison with the 617 MHz image 
in Fig.~\ref{fig:scheme}(b), in all panels we report the lowest contour at 617 MHz in magenta. 

The central part of the minihalo is bright at all frequencies, with a size 
of $r\sim 30^{\prime\prime}$ ($\sim 80$ kpc). The much fainter tail, which is best 
detected at 617 MHz (Fig.~\ref{fig:scheme}), 
is also visible at 317 MHz (a) and 1.48 GHz (c). Only peaks of its emission are 
visible at the $3\sigma$ level in the other panels. Its maximum length
is $\sim 1^{\prime}.4$ ($\sim230$ kpc) at 617 MHz.

\begin{figure}
\centering
\includegraphics[angle=0,width=8cm]{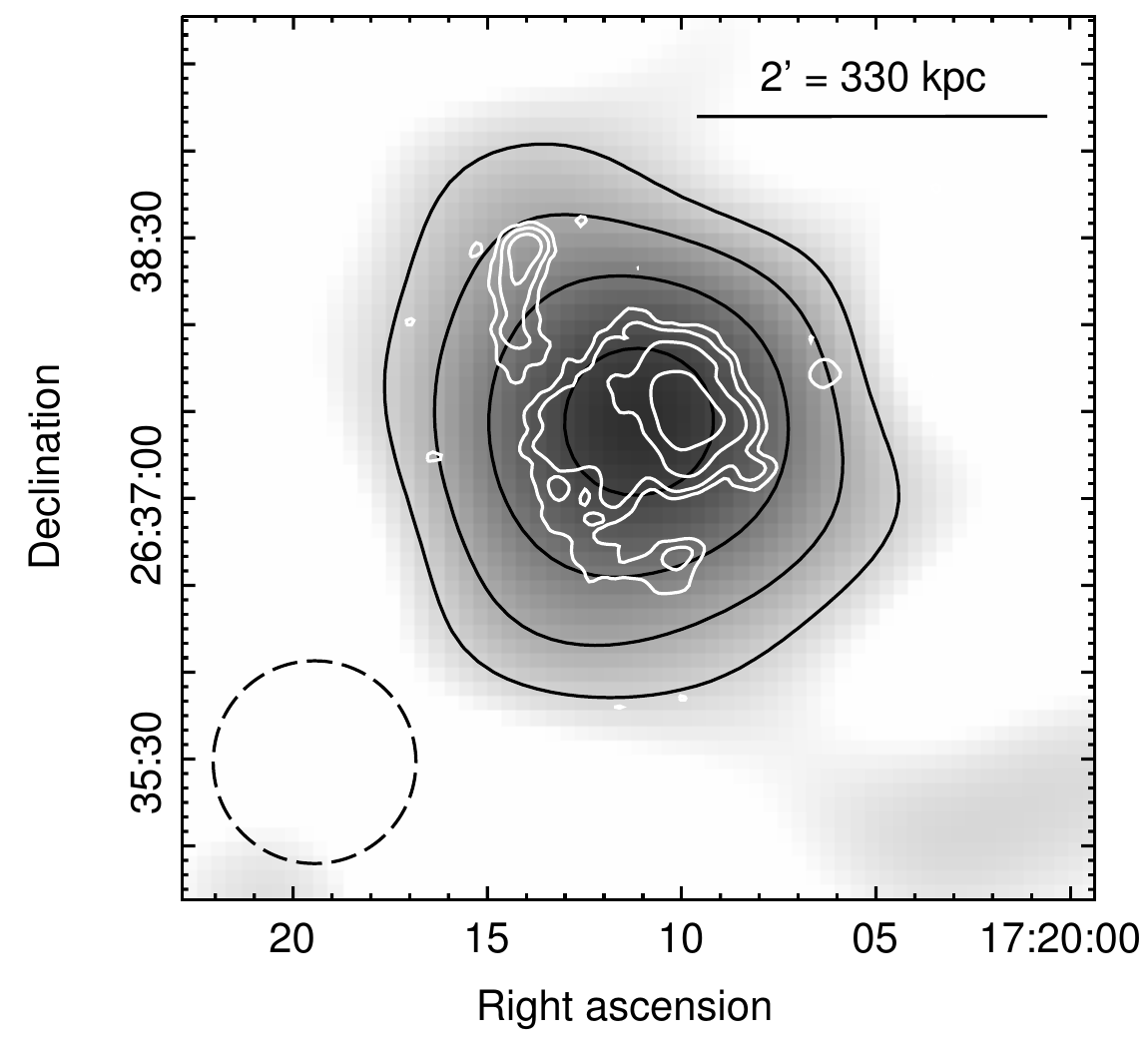}
\smallskip
\caption{VLSS-Redux image at 74 MHz (grayscale and black contours)
of the cluster central region. The restoring beam (dashed circle) 
is $75^{\prime\prime}$ and local rms noise level is 68 mJy beam$^{-1}$. 
Black contours start
at 200 mJy beam$^{-1}$ and then scale by a factor of 2. Radio contours
at 617 MHz at 0.09, 0.36, 1.4 and 5.8 mJy beam$^{-1}$ are reported in white
(from Fig.~\ref{fig:scheme}).}
\label{fig:vlss}
\end{figure}

We retrieved an image at 74 MHz from the VLSS-Redux\footnote{{\em VLA} Low-Frequency Sky Survey 
Redux (Lane er al. 2012).}, which we present in Fig.~\ref{fig:vlss} as grayscale and black 
contours. The angular resolution of the image is $75^{\prime\prime}$. A bright, marginally extended 
source with $3.1\pm0.3$ Jy\footnote{This value is in the \cite{2012MNRAS.423L..30S} 
flux density scale.} is detected. A comparison with the higher-resolution image at 617 MHz 
(white contours) indicates that the VLSS-Redux source is a blend of the minihalo and head-tail 
emissions. After subtraction of the expected flux densities of the BCG 
and head tail at this frequency ($\sim 143$ mJy and $\sim 80$ mJy, based on
the spectral indices in Table 3), a residual flux of $\sim 2.9$ Jy
is estimated for the minihalo.

In Table 4, we summarize the properties of the minihalo and 
of its components. The flux densities were measured on Figs.~\ref{fig:scheme} 
and \ref{fig:mh_full} using the task TVSTAT and the magenta and blue regions 
shown in Fig.~\ref{fig:scheme}. All fluxes are point-source subtracted. The
errors on the minihalo flux density $S_{\rm MH}$ were estimated as 

\begin{equation}
\sigma_{S_{\rm MH}} = \sqrt{(\sigma_{\rm cal} S_{\rm MH})^2 + (rms \sqrt{N_{\rm beam}})^2 + \sigma_{\rm sub}^2}
\end{equation}

\noindent which takes into account the uncertainty on the flux 
density scale ($\sigma_{\rm cal}$), the image rms level weighted by the 
number of beams in the minihalo region ($N_{\rm beam}$), and the uncertainty 
$\sigma_{\rm sub}$ in the subtraction of the central radio galaxy  
from the total flux density measured in the image \citep{2013ApJ...777..141C}. 
This latter was estimated as 

\begin{equation}
\sigma_{sub}^2= (I_{MH,s} \times A_{s})^2
\end{equation}

\noindent where $A_{s}$ is the area occupied by the radio galaxy and 
$I_{MH,s}$ is the average surface brightness of the minihalo in the proximity 
of the source.

A proper comparison of the minihalo properties at different frequencies 
should be based on images obtained selecting the same interval of projected 
baselines. The minihalo emission spans an area of $<2^{\prime}$ in diameter, 
which is sampled by baselines longer than 1 k$\lambda$. As clear from Table 2, 
all data sets used here have a minimum projected baseline of $\sim 1$ $k\lambda$ 
or shorter, ensuring a proper detection of the entire minihalo emission (but 
see \S\ref{sec:mh_inj} for limitations of the short observations at 4.86 GHz and 8.44 GHz). 
We thus selected a common $u-v$ range of 1-50 k$\lambda$ and made images 
(using natural weights) at all frequencies to measure the flux densities 
of the minihalo and of its components and compare them with the values 
measured on the naturally-weighted images in Fig.~2 (Table 4). All flux densites were 
found to be consistent within the errors with the values in Table 4.
We also measured the minihalo flux density on uniformly-weighted images 
and found consistent values with those in Table 4 at 317, 617, 1280 and 1480 MHz.
For the flux density measurements at 4.86 GHz and 8.44 GHz see section 5.2, 
where a careful analysis of the possible missing flux is presented.

It is clear that the central part accounts for most of the minihalo 
flux density at all frequencies, ranging from $\sim 80\%$ at 317 MHz to $\sim 90\%$ at 
the highest frequencies. At 8.44 GHz, only an upper limit on the flux density of the tail 
can be placed; the existing observation lacks both the sensitivity and short-baseline 
coverage to reliably detect this faint and very extended region of the minihalo 
(see \S\ref{sec:mh_inj}).

%
%
%
%

%
%
%
%

\section{Radio spectral analysis}

\subsection{Radio spectra of the minihalo and BCG}
\label{sec:spectrum_bcg}

\begin{figure*}
\centering
\includegraphics[angle=0,width=8cm]{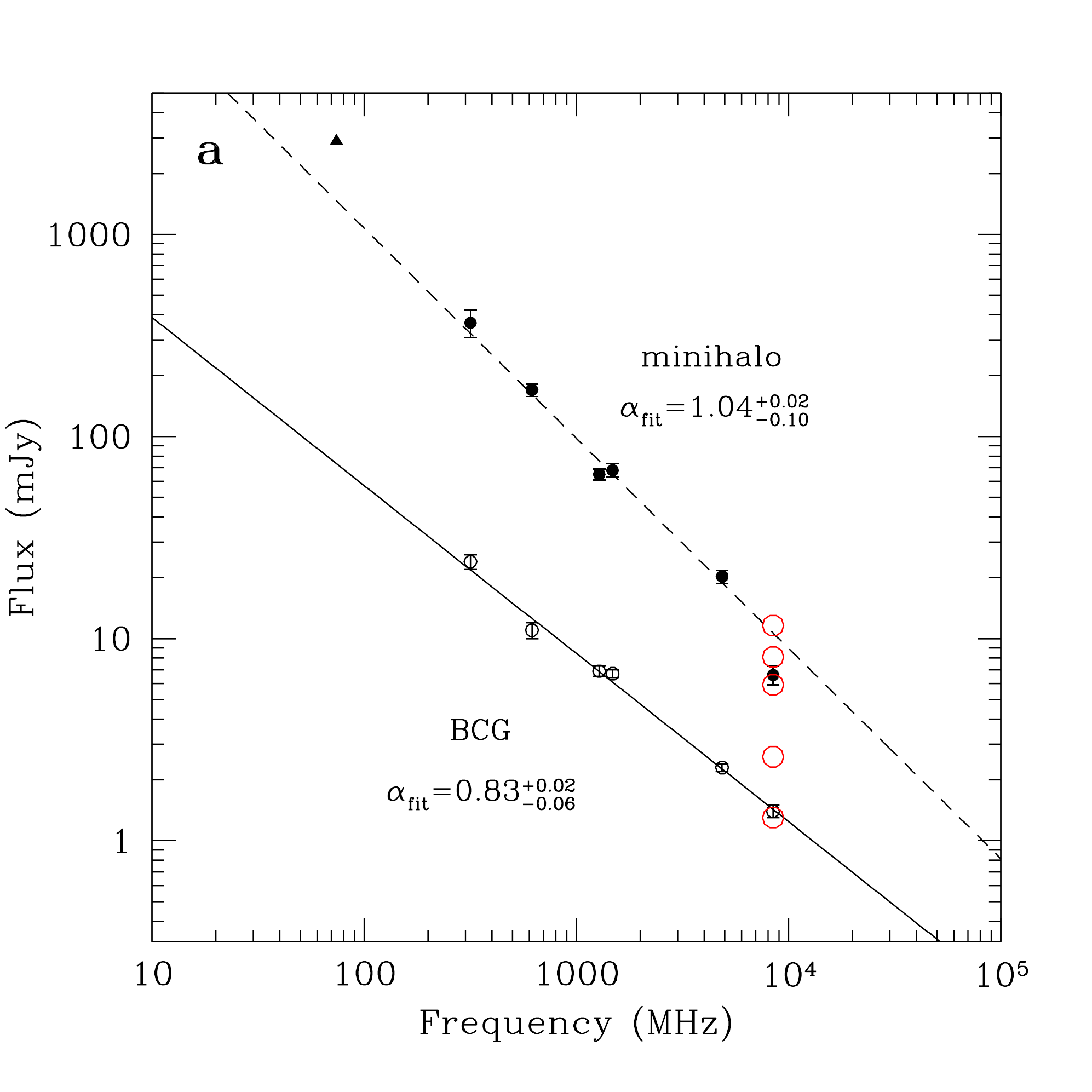}
\includegraphics[angle=0,width=8cm]{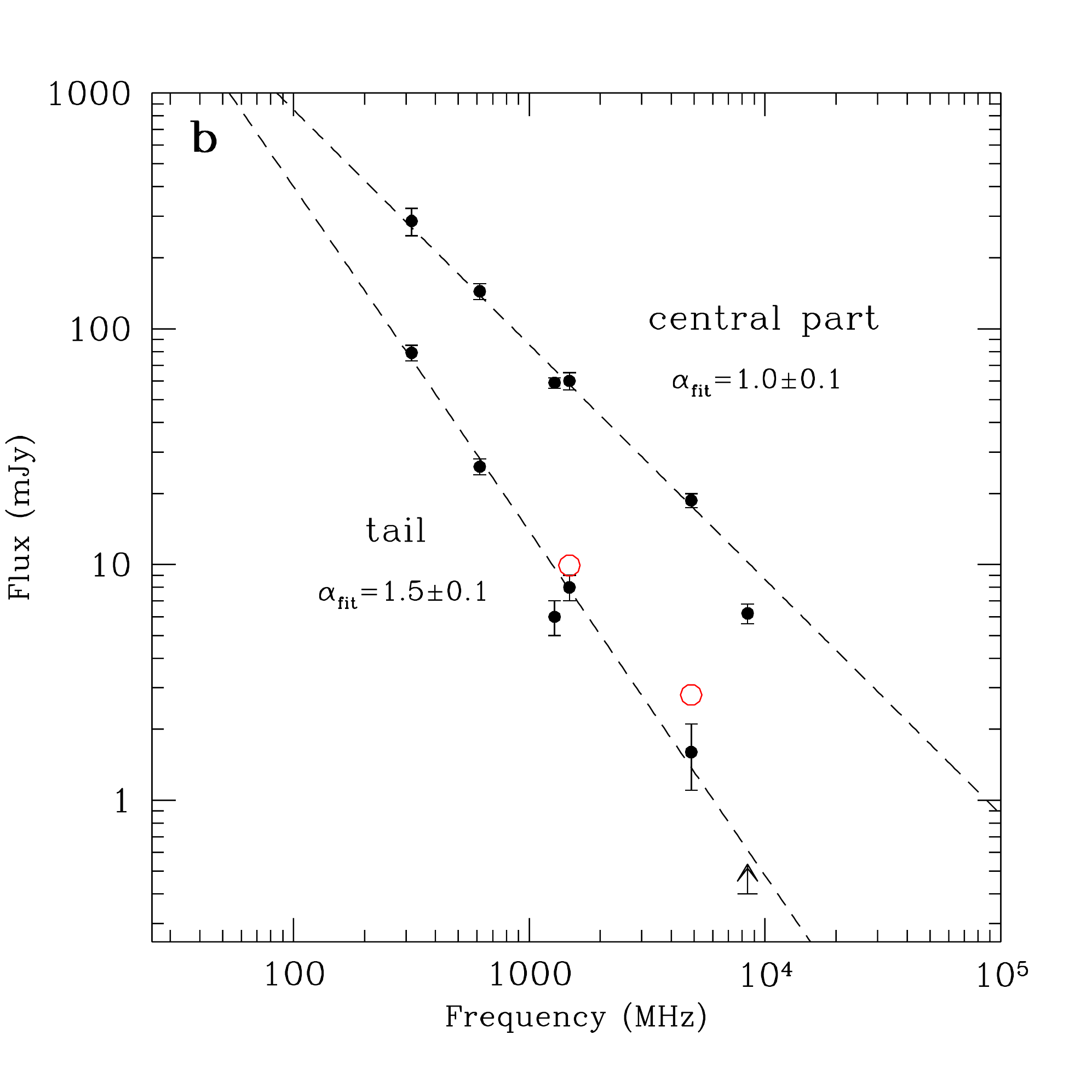}
\smallskip
\caption{(a) Radio spectra of the BCG (empty circles)
and minihalo (filled circles) between 317 MHz and 8.44 GHz. 
The solid line is a power-law fit to the BCG spectrum.
The dashed line is a power-law fit to the minihalo spectrum
between 317 MHz and 4.86 GHz. The slopes provided by the 
fits are reported. The filled triangle is the flux density 
of the minihalo at 74 MHz, estimated from the VLSS-Redux image (see \S \ref{sec:mh}).
Red empty circles are the recovered flux densities 
of a minihalo model based on the 617 MHz image and injected 
into the 8.44 GHz data set assuming $\alpha=$1.0, 1.1, 1.2, 1.4 and 1.5 (from top to bottom).
(b) Radio spectra of the minihalo
components between 317 MHz and 8.44 GHz. The dashed lines are power-law fits 
to the data between 317 MHz and 4.86 GHz. The slopes provided by the 
fits are reported. Red empty circles are the 
recovered flux densities in the region of the tail 
of a minihalo model based on the 617 MHz image and injected 
into the 1.48 GHz and 4.86 GHz data sets assuming $\alpha=1$.}
\label{fig:mh_sp}
\end{figure*}

In  Fig.~\ref{fig:mh_sp}(a), we show the radio spectra of the BCG (empty circles) and minihalo 
(filled circles) between 317 MHz and 8.44 GHz, based on the flux densities in Tables 
3 and 4.

The BCG spectrum can be fitted by a single power-law model over the entire frequency range,
with a slope $\alpha_{\rm fit}=0.83$, typical of an active radio galaxy \citep[e.g.,][]{1992ARA&A..30..575C}.

Unlike the BCG, the minihalo spectrum is well described by a single power law 
(with $\alpha_{\rm fit}=1.04$) only up to $\sim 5$ GHz; above this frequency, 
the spectral index steeepens to $\alpha\approx 2$. The filled triangle is 
the 74 MHz flux density measured on the VLSS-Redux image ($\S\ref{sec:mh}$). 
Despite the uncertainties in the subtraction 
of the BCG and head-tail emissions, the 74 MHz estimate appears in 
reasonable agreement with the higher-frequency data points.

In Fig.~\ref{fig:mh_sp}(b), we show the spectra of the minihalo components. 
Up to 4.86 GHz, both components have a power-law spectrum with
$\alpha_{\rm fit}=1.0$ and $\alpha_{\rm fit}=1.5$, respectively. As seen 
for the entire minihalo, the spectrum of the central part steepens 
above 5 GHz; a possible steepening is also visible for the tail.

%
%
%

%
%
\begin{figure*}
\centering
\includegraphics[angle=0,width=16cm]{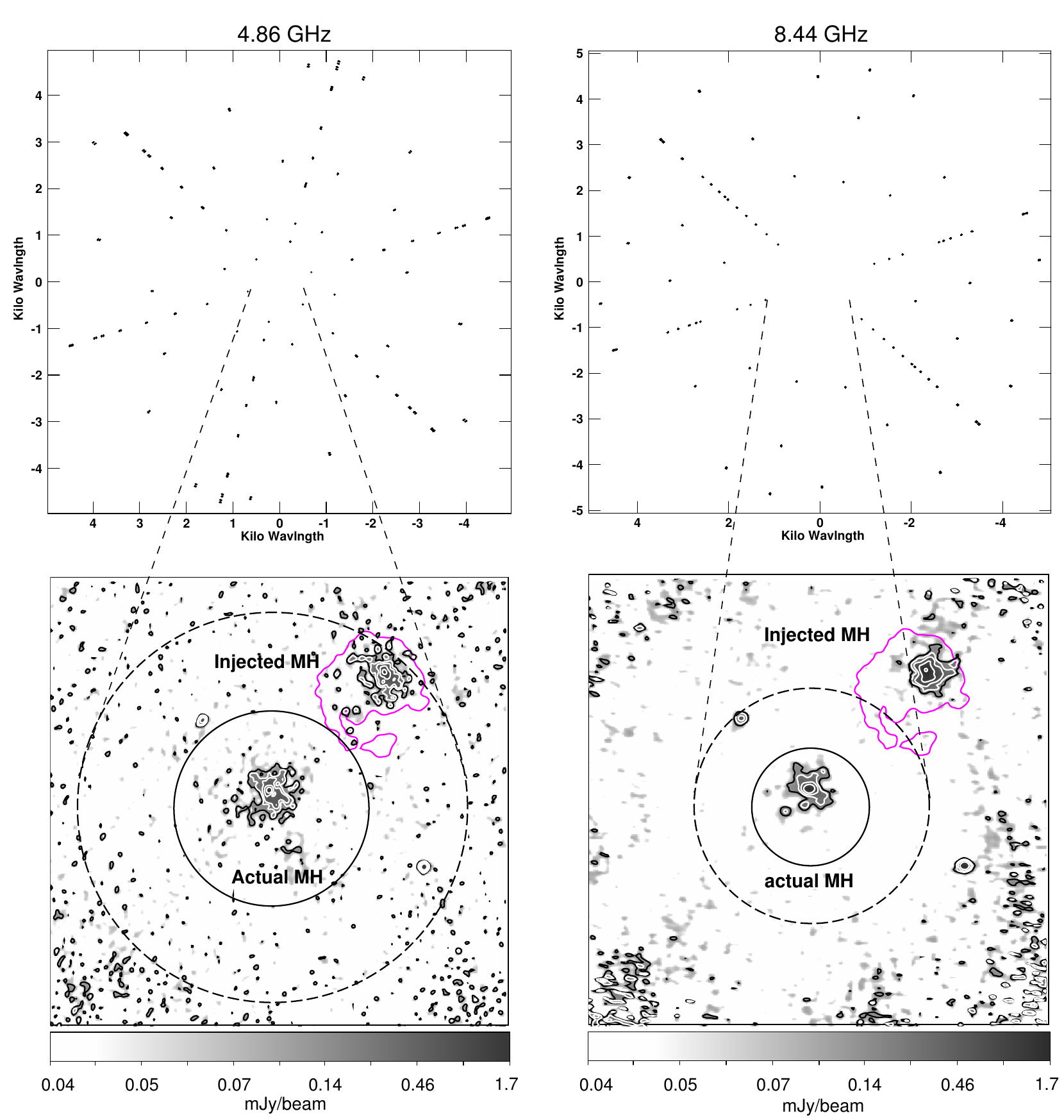}
\smallskip
\caption{{\em Upper panels:} inner portion of the $u-v$ plane sampled by 
the VLA data sets at 4.86 GHz (C configuration) and 8.44 GHz. 
{\em Lower panels:} examples of minihalo injections 
in the VLA data sets at 4.86 GHz and 8.44 GHz. A minihalo model 
based on Fig.~\ref{fig:scheme} (the lowest contour is reported here in magenta) 
has been injected at $\sim 2^{\prime}$ from the phase center and then imaged, along 
with the actual minihalo, using natural weighting. Contours are spaced 
by a factor of 2 starting from 0.1 mJy beam$^{-1}$. The radio beams are as those in 
Figs.~\ref{fig:mh_full}(e) and ~\ref{fig:mh_full}(f). The images have been corrected 
for the primary beam attenuation. In both images, the solid circle shows the 
largest angular scales which can be reliably imaged (the diameter is $2.5^{\prime}$ 
at 4.86 GHz and $1.5^{\prime}$ at 8.44 GHz). For comparison, the scale detectable by 
longer observations in the same array configurations ($5^{\prime}$ and $3^{\prime}$)
are reported as dashed circles.
}
\label{fig:mh_inj}
\end{figure*}
%
%

\subsection{Minihalo injections and flux density losses at 4.86 GHz and 8.44 GHz}\label{sec:mh_inj}

A proper determination of the flux density of an extended
and low surface brightness source such as a minihalo 
requires high sensitivity and good sampling 
of the short $u-v$ spacings, which provide information on the 
large angular scales. In the minihalo spectrum in Fig.~\ref{fig:mh_sp}, 
all data points at $\le 1.48$ GHz derive from deep, pointed 
observations with good coverage at short baselines and suited 
to image extended emission on $\ge 2^{\prime}$ scales (Table 2; 
to be compared to the angular size of the minihalo, which 
is $\sim 1^{\prime}.6$ in diameter).

The observations at 4.86 GHz and 8.44 GHz are instead only few minutes long 
and their $u-v$ coverage is therefore much poorer. In Fig.~\ref{fig:mh_inj} 
(upper panels), we show the inner portion of the $u-v$ planes at 4.86 GHz 
(C configuration) and 8.44 GHz. It is clear that the data sampling is very sparse, 
especially at 8.44 GHz, resulting into a possible underestimate of the 
diffuse emission at these frequencies. The maximum angular structure 
that can be reliably imaged ($\theta_{\rm LAS}$) is also affected by 
the poorer $u-v$ coverage. VLA full-synthesis observations at 4.9 
GHz and 8.4 GHz (in these same configurations) should be able to detect
emission on maximum scales of $\sim 5^{\prime}$ and $\sim 3^{\prime}$, respectively
(dashed circles in the lower panels of Fig.~\ref{fig:mh_inj}). However, for shorter 
observations like the ones used in this paper, $\theta_{\rm LAS}$ is considerably smaller, 
i.e., $\lax 2^{\prime}.5$ and $\lax 1^{\prime}.5$ respectively\footnote{http://science.nrao.edu/facilities/vla/proposing/oss/ossjan09.pdf} (solid circles). 
This implies that, while the 4.86 GHz observation is potentially able to image the 
whole minihalo, its largest-scale emission -- the tail -- cannot be fully detected 
at 8.44 GHz and, therefore, the flux measured for such component at this frequency 
must be considered as a lower limit (Table 4).

To evaluate possible losses of the minihalo emission at 4.86 GHz and 8.44 GHz, we have 
adopted a procedure similar to the injection of {\em fake} giant radio halos 
in GMRT data sets by \cite{2007ApJ...670L...5B}, \cite{2008A&A...484..327V} and
\cite{2013A&A...557A..99K}. We injected 
a minihalo model in the $u-v$ data at 4.86 GHz and 8.44 GHz and then imaged both data 
sets with the same parameters we used to produce the actual-minihalo images in Figs.~\ref{fig:mh_full}(e) and (f). 
We then measured the flux densities of the {\em fake} minihalos and compared them to the injected ones, 
obtaining an estimate of the losses. 

For their statistical purposes, \cite{2013A&A...557A..99K} and previous two works 
injected an ``average'' fake radio halo, composed of a set of optically thin concentric 
spheres with different radius and flux 
density. For the minihalo, because of its asymmetric morphology, 
we chose to inject a model based on the observed surface brightness distribution 
rather than using a set of spheres which would not describe well the spiral-shape 
structure of the minihalo. Our results will depend on the morphology of the 
injected minihalo, consequently we used our highest-quality image -- the 
\gmrt\ image at 617 MHz (Fig.~\ref{fig:scheme}) -- as initial model.

We extracted the CLEAN components of the minihalo from the 617 MHz image and scaled them 
at 4.86 GHz and 8.44 GHz assuming $\alpha=1$, which is similar to the spectral index 
of the minihalo between 317 MHz and 1.4 GHz (Table 4). We then injected 
the components in the 4.86 GHz and 8.44 GHz data sets using 
the task UVSUB in AIPS, selecting a region free of sources and as near as possible 
to the phase center to minimize primary-beam attenuation. We ran a number of injections 
using different regions; an example is shown in Fig.~\ref{fig:mh_inj} (lower panels), 
where we injected a fake minihalo $\sim 2^{\prime}$ north-west of the actual minihalo. 
We found a good agreement between the morphology of the injected minihalo and the 
observed ones. More than $95\%$ of the flux density injected at 4.86 GHz is recovered 
in the images, indicating that possible flux losses are contained within the errors on the 
minihalo flux density at this frequency (Table 4). We may expect even higher 
values of recovered emission, if the minihalo were injected at the phase center (where the 
actual minihalo is located), where the sensitivity of the observation is the highest.

At 8.44 GHz, we estimated a flux density loss of $\sim 7\%$, thus still 
within the flux density uncertainty in Table 4. However, the amount of flux 
density recovered in the area of the fake minihalo will depend on the initial total flux 
density injected in that region, and thus on the $\alpha$ used to scale the 617 MHz flux 
density to 8.44 GHz. To evaluate this effect, we injected fake minihalos steepening 
progressively the spectral index from $\alpha=1$ to 1.5 (red empty circles in Fig.~6a). 
We found that an increasing fraction of the injected flux density is lost when the injected 
flux decreases, with losses of $\sim 10-15\%$ for $\alpha=1.1$, $\sim 20\%$ for 
$\alpha=1.2$, $\sim 40\%$ for $\alpha=1.4$, and exceeding $50\%$ for $\alpha=$1.5.

We also evaluated the fraction of flux density that is lost for the minihalo components individually. The 617 MHz CLEAN 
components were scaled to the higher frequencies using $\alpha=1$ for the central part 
and $\alpha=1.5$ for the tail (Table 4). We found that possible losses 
at 4.86 GHz are within the flux density uncertainties of $\sim 7\%$ and $\sim 30\%$ 
for the central region and tail, respectively (Table 4).
At 8.44 GHz, we recovered $\sim 93\%$ of the emission of the central part (thus again 
within the error), but we are not able to recover the tail. This is expected given 
the $u-v$ coverage limitations at this frequency described above. 

As a further test to determine whether the spectrum of the tail is intrinsically 
steeper than that of the central part (Fig.~\ref{fig:mh_sp}b), or if the observed 
steepness is affected by flux density losses, we injected the tail in the 1.48 GHz 
and 4.86 GHz data sets assuming that it has the same $\alpha=1$ as the central part. 
The empty red circles in Fig.~\ref{fig:mh_sp}b show the flux densities recovered by the
imaging at these frequencies. No significant losses are found at 1.48 GHz and less 
than $\sim 15\%$ loss is estimated at 4.86 GHz. This indicates that the observed 
steepness is not driven by flux losses at these frequencies.

We conclude that the high-frequency steepening observed in the minihalo spectrum 
(Fig.~\ref{fig:mh_sp}) is most likely real. This is also true for its central part, 
which dominates the minihalo emission, whereas the steepening in the spectrum of 
the tail, which does not contribute significantly to the total flux of the minihalo, 
may be caused by an underestimate of its flux density at 8.44 GHz. However, due to possibly 
significant flux density losses at 8.44 GHz, we are unable with the current data to 
determine the actual change of slope at high frequencies.

\subsection{Study of the spectral index distribution}
\label{sec:index_map}

\begin{figure*}
\centering
\hspace{1.5cm}\includegraphics[angle=0,width=15.5cm]{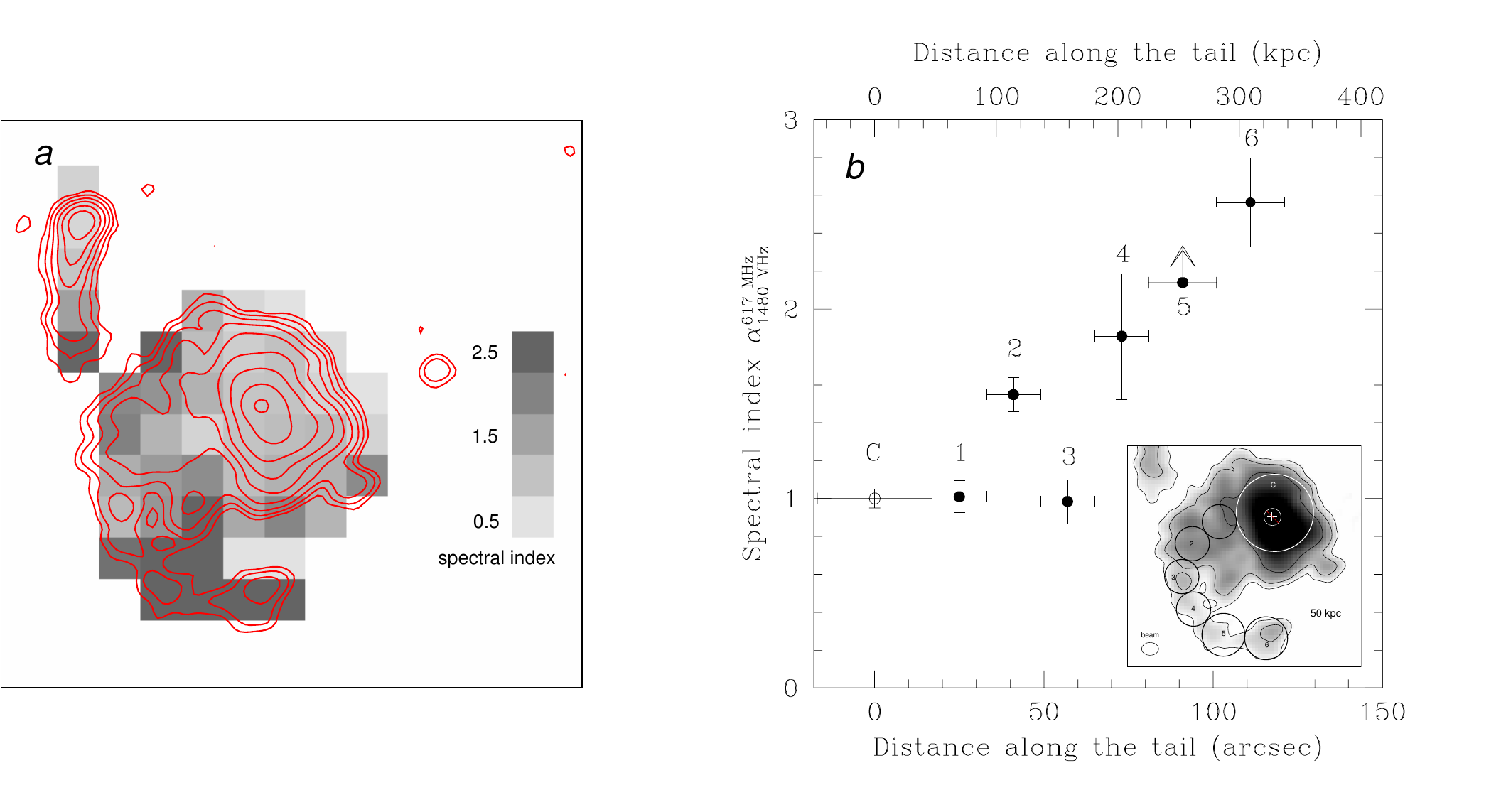}
\smallskip
\caption{(a) Grayscale image of the spectral index distribution between 617 MHz
and 1480 MHz in the minihalo and head-tail radio galaxy. The image has been computed 
from images with similar noise (30 $\mu$Jy beam$^{-1}$) and same $u-v$ range and restoring 
beam of $8^{\prime\prime}\times6^{\prime\prime}$. Overlaid are the 617 MHz contours from Fig.~2a.
(b) Spectral index between 617 MHz and 1480 MHz as a function of the distance from the cluster 
center (white cross) along the minihalo tail (black, filled points). The profile has been derived 
using the independent circular regions shown in the inset, with $r=8^{\prime\prime}$ (regions 1 to 4)
and $r=10^{\prime\prime}$ (regions 5 and 6). For a comparison, we also report 
the spectral index of the central part of the minihalo (C) 
computed in a $r=18^{\prime\prime}$ region (white circle in the inset) and
excluding the point source at the BCG. 
Errors are $1\sigma$. The ellipse in 
the lower left corner of the inset shows the beam size.}
\label{fig:spix}
\end{figure*}

We obtained a spectral index map of the minihalo by comparing a pair of images at 617 MHz 
and 1.48 GHz. We selected these frequencies because of the similar sensitivity 
($\sim 30$ $\mu$Jy beam$^{-1}$) and $u-v$ coverage of the observations, which ensure a 
good description of the minihalo flux density and morphology. We produced images
at 617 MHz and 1.48 GHz imposing the same $u-v$ range and restoring beam of 
$8^{\prime\prime}\times6^{\prime\prime}$, and corrected them for the primary-beam attenuation. 
The two images were then aligned and binned by 8 pixels ($12^{\prime\prime}$). Bins 
with large uncertainties on the spectral index were blanked.
The resulting qualitative spectral index image is shown in Fig.~\ref{fig:spix}(a), 
with the 617 MHz contours overlaid to provide a reference for the source morphology.

The spectral index distribution in central region of the minihalo  
is quite uniform with an average $\alpha\sim 1$, in good agreement with 
the slope of the total spectrum in this frequency range 
(Fig.~\ref{fig:mh_sp}(b)). The spectral index of the tail is 
steeper and it increases systematically from $\alpha\sim 1$ to
$\alpha \sim 2.5$ with increasing distance from the center. 
To analyzed such trend, we first extracted the flux densities at 617 MHz and 1.48 GHz 
in circular regions along the tail, as shown by the black circles in the inset of Fig.~\ref{fig:spix}(b), 
and then computed the corresponding spectral indices. The size of each circle 
was chosen to be larger than one beam to sample independent regions 
($r=8^{\prime\prime}$ for regions 1 to 4, and $r=10^{\prime\prime}$ for regions 5 and 6). 
In Fig.~\ref{fig:spix}(b), we show the spectral index as a function of the distance 
from the cluster center along the minihalo tail. 
For a comparison, we also report the spectral index of the 
central region computed in a $r=18^{\prime\prime}$ region, as shown 
by the white circle in the inset. The profile is consistent with
the trend seen in Fig.~\ref{fig:spix}(a); the increase in spectral 
index is $\Delta\alpha\sim 1.6$ along the 230 kpc-long tail.

Fig.~\ref{fig:spix}(a) also shows the spectral index distribution in 
the head tail radio source. As commonly observed in this type of radio galaxies, 
a gradual increase of $\alpha$ is visible going from the region of the head, 
which has the flattest spectral index ($\alpha\sim 0.5$), to the end of the tail. 
Such steepening reflects the progressive aging of the relativistic electrons as 
they travel away from the core \citep[e.g.,][]{2010A&A...514A..76M}.

%
%
%
%
%

\section{Discussion}
\label{sec:disc}

\subsection{Physical origin of the minihalo}
\label{sec:origin}

Radio spectral properties of minihalos --- the spectral shape and its
spatial variations --- provide a discriminator for the physical mechanism
responsible for the generation of the radio-emitting relativistic electrons
in cluster cores (see Brunetti \& Jones 2014 for a review).  However,
because of the rarity of minihalos and technical difficulty of separating
the central bright radio source from an often much fainter extended 
source that is the minihalo, spectral data on the minihalos 
are scarce.  Until the present work, only three minihalos had known 
spectra, none of them detailed. Perseus (Sijbring 1993) and Ophiuchus 
\citep{2010A&A...514A..76M} each have measurements at three frequencies, which 
are consistent with power-law spectra (although there is a hint of a 
high-frequency steepening in Ophiuchus). The recently discovered minihalo 
in RX\,J1532.9+3021 \citep{2013ApJ...777..163H,2014ApJ...781....9G}
has measurements at four frequencies, which are again consistent 
with a power law up to 1.4 GHz, while hinting at steepening at higher 
frequencies \citep{2014ApJ...781....9G}. Spatially resolved measurements 
of the minihalo spectral slope were reported for Perseus (Sijbring 1993) 
and Ophiuchus \citep{2010A&A...514A..76M}, but their quality and spatial coverage 
at the frequencies used for the spectral index mapping is limited.

In this paper, we presented a new high-sensitivity, well-resolved,
multifrequency dataset for the radio minihalo in RXJ1720.1, which
makes it the best-studied object of this class. We report the 
detection of a possible high-frequency break in the spectrum 
of this minihalo (\S\ref{sec:spectrum_bcg}), though deeper high-frequency 
observations are needed to quantify the change of the slope. 
Furthermore, we detected significant, systematic spatial variations of 
the minihalo spectrum --- the spectral index steepens with the distance from 
the cluster (and minihalo) center
(\S\ref{sec:index_map}).

\begin{figure*}
\centering
\includegraphics[angle=0,width=16cm]{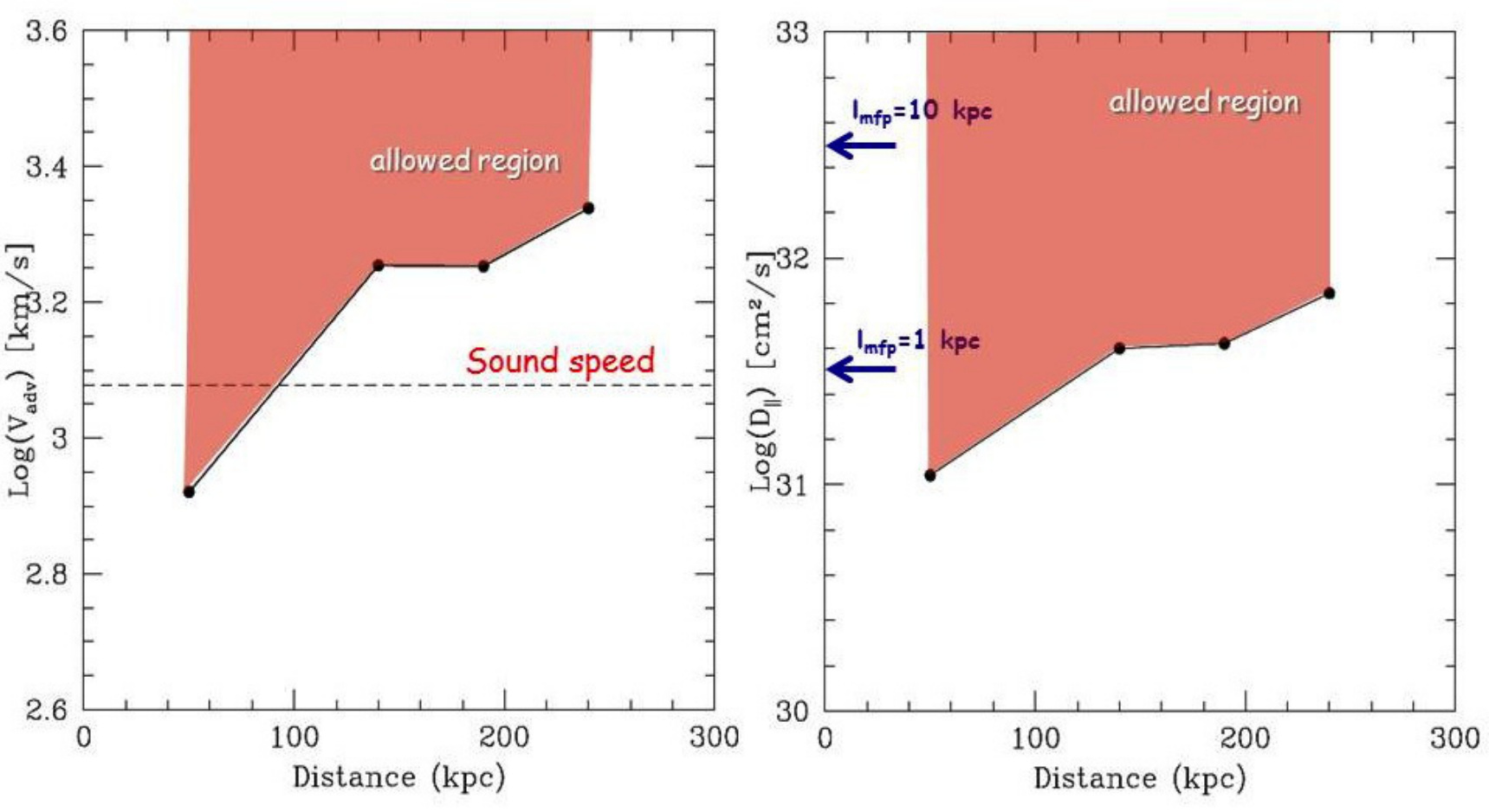}
\smallskip
\caption{{\em Left:} Minimum required advection velocity as a function of the 
distance along the tail from region 1 in Fig.~8b. The dashed line indicates 
the value of the local sound speed. {\em Right:} 
Minimum required diffusion coefficient along the magnetic field lines 
as a function of the same distance. We also show the effective particle 
mean free path for two relevant values of the diffusion coefficient (see text).
In both panels, region 3 is omitted.}
\label{fig:age}
\end{figure*}

A high-frequency break in the total spectrum favors a reacceleration scenario 
over plausible alternatives, such as the pure ``secondary'' model, in which 
the radio-emitting relativistic electrons are produced by hadronic collisions 
and the spectrum of the electrons is expected to be a power law 
\citep{2004A&A...413...17P}. 
The spectrum of a minihalo powered by turbulence is
expected to be a power law with a break or cutoff, which is a consequence of the 
low efficiency of turbulence as an acceleration mechanism
\citep[e.g.,][]{2002A&A...386..456G}. The balance between the energy gain
through reacceleration and the losses (primarily via synchrotron and inverse Compton (iC)
radiation) sets a cutoff in the resulting energy distribution of the
electrons that, in turn, creates a break in the synchrotron spectrum at a
frequency determined by the acceleration efficiency (see, e.g., simulations
of this process in a realistic turbulent cluster core by Z13).  
Our observed spectrum of the RX\,J1720.1 minihalo exhibits a possible 
steepening of the spectral index above 5 GHz. We see indication of a 
high-frequency spectral break in the spectrum of the central part (the dominant contributor to the total
flux) and, separately, a hint of steepening in the tail.
The significant steepening of the spectral index with increasing distance
along the spiral tail (Fig.~\ref{fig:age}b) further favors 
a reacceleration scenario, because such strong spectral variations 
across the minihalo are not expected in the ``secondary'' model (ZuHone et al. 2014).
A similar steepening of the radio
spectrum with increased distance from the cluster center has been reported
by \cite{2010A&A...514A..76M} for the Ophiuchus minihalo (although for that
cluster, only a fraction of the minihalo was detected at both frequencies
that were used for the spectral map). Spatial variations of the spectral
slope were also reported for the Perseus minihalo (Sijbring 1993); in fact,
spatial variations in Perseus prompted the hypothesis of turbulent
reacceleration for the origin of the minihalos \citep{2002A&A...386..456G}.

Because of their smaller size, the requirement of in-situ acceleration or injection
of the radio-emitting electrons in minihalos is less strict than in giant radio halos. 
For this reason, here we also discuss the possibility 
that the observed spectral steepening along the spiral tail (Fig.~\ref{fig:spix}) is simply 
caused by aging of relativistic electrons that originate in the central region of the minihalo
(e.g., in the radio galaxy) and are advected, or diffuse, to its periphery.
In this case, steeper spectra seen in Fig.~\ref{fig:spix} are simply
produced by older electrons. To evaluate whether advection is a possibility,
we can estimate the age of the relativistic electrons and check what gas
velocity is required to transport them to the observed distances. Since we
do not know the magnetic field in this cluster, we will use the field
strength for which the total synchrotron and iC energy losses are at their
minimum, $B \simeq B_{iC}/\sqrt{3} \sim 2.5 \, \mu$G, where $B_{iC} = 3.2
(1+z)^2\,\mu$G is the field whose energy density equals that of the Cosmic
Microwave Background (CMB). For the fields above or below this value, the
age of the electrons would be shorter, so we assume this value for a
conservative estimate. Fig.~\ref{fig:age}(a) shows the minimum advection
velocity for this field that would match the observed spectral steepening of
the minihalo, as a function of distance along the spiral tail. Here we
assume an injection (initial) spectral power-law slope of 1, consistent with
the spectrum of the central region of the minihalo. The resulting velocities
are several times larger than the typical gas velocities found in numerical
simulations of sloshing cores (e.g., Ascasibar \& Markevitch 2006; ZuHone et
al. 2011, Z13), and higher than even the sound speed. Thus, the advection
explanation is not feasible.

Particle diffusion along the magnetic field lines may also play a role. In
this case, the maximum age of the relativistic electrons determines the
minimum spatial diffusion coefficient along the field lines for GeV
particles, $D_{\Vert} \simeq 1/4 L^2 / \tau$, where $\tau$\/ is the
diffusion time and $L$\/ is the diffusion scale. Again, for a conservative
estimate, we assume an optimistic picture in which a magnetic field with
intensity $B \sim 2.5 \, \mu$G (that maximizes the electron lifetime) is
mostly aligned along the tail of the minihalo, as seen in MHD simulations of
the sloshing cool cores \cite{2011ApJ...743...16Z}. We also assume that there are
no significant perturbations or waves on small scales that would reduce the
diffusion along the field lines due to scatter of the particle pitch angle.
The spatial diffusion coefficient required to explain the observed spectral
steepening is shown in Fig.~\ref{fig:age}(b). The required values are very
large --- for example, orders of magnitude higher than current estimates for our Galaxy
\citep{1990acr..book.....B}.

Even in the absence of micro-scale perturbations that could strongly reduce 
diffusion along the
field lines, the field should be advected and perturbed by large-scale gas
motions, including turbulence. The required minimum values of $D_{\Vert}$
derived in Fig.~\ref{fig:age}(b) place a lower limit on the effective
mean-free path of particles and --- because particles travel strictly along
the field lines --- on the minimum coherence (or tangling) scales of the
magnetic fields.  Using $D_{\parallel} \sim 1/3 c l_{\rm mfp}$ from Fig.~\ref{fig:age}(b),
where $c$ is the speed of light, we find $l_{\rm mfp} > 5$ kpc, 
which is in tension with the minimum scales of magnetic field fluctuations
observed in similar environments \citep{2011A&A...529A..13K}. Thus, 
diffusion of relativistic electrons originating in the central region 
outwards along the field lines seems to be unable to explain the observed 
spectral behavior in the minihalo tail, although this possibility cannot 
be firmly ruled out due to our very scarse knowledge of the magnetic field properties
and particle/diffusion transport in these environments.

%
%

\subsection{Minihalo confinement}

\begin{figure*}
\centering
\hspace{-1.5cm}
\includegraphics[width=7.1cm]{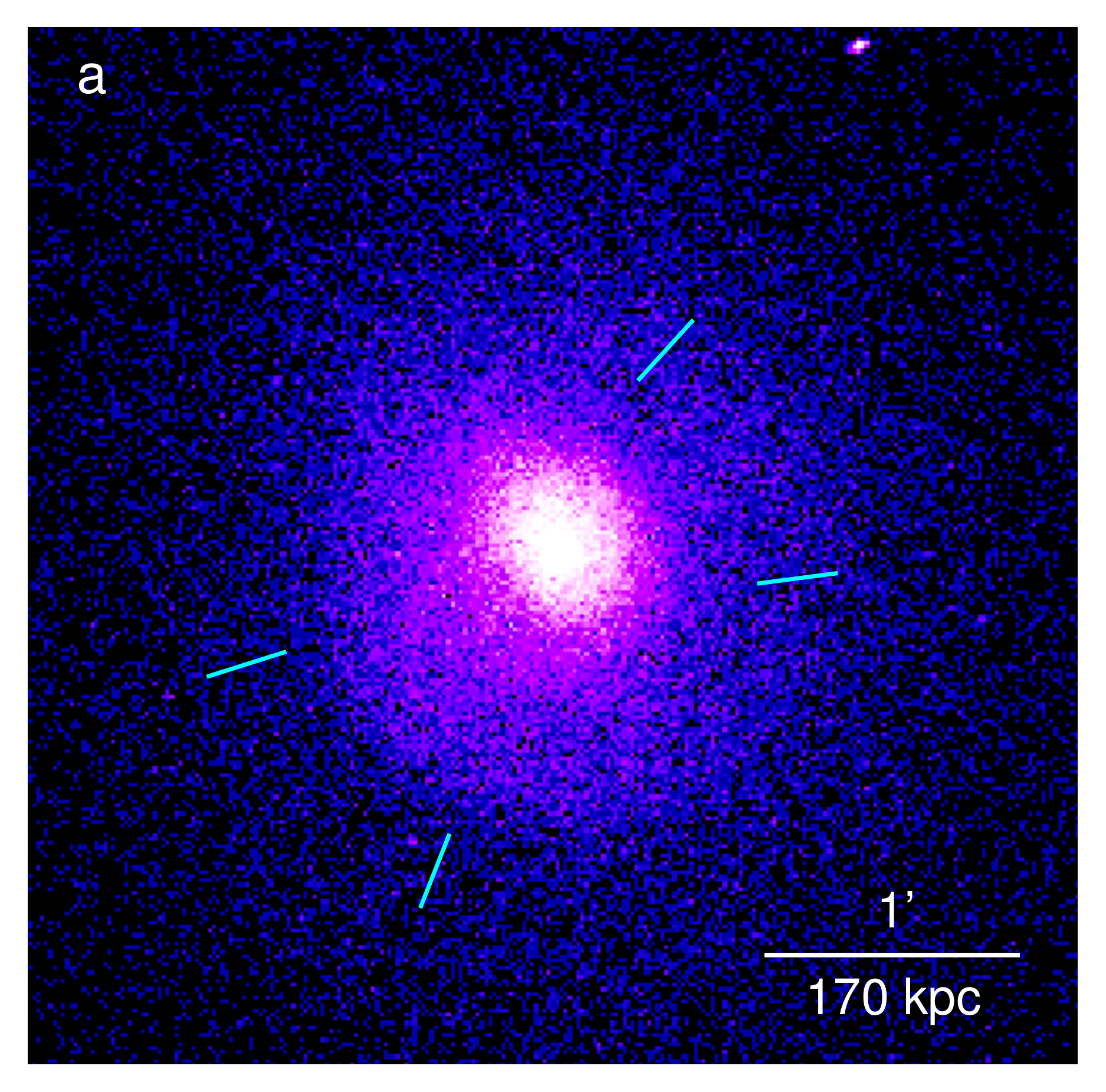}
\hspace{0.5cm}
\includegraphics[width=7.1cm]{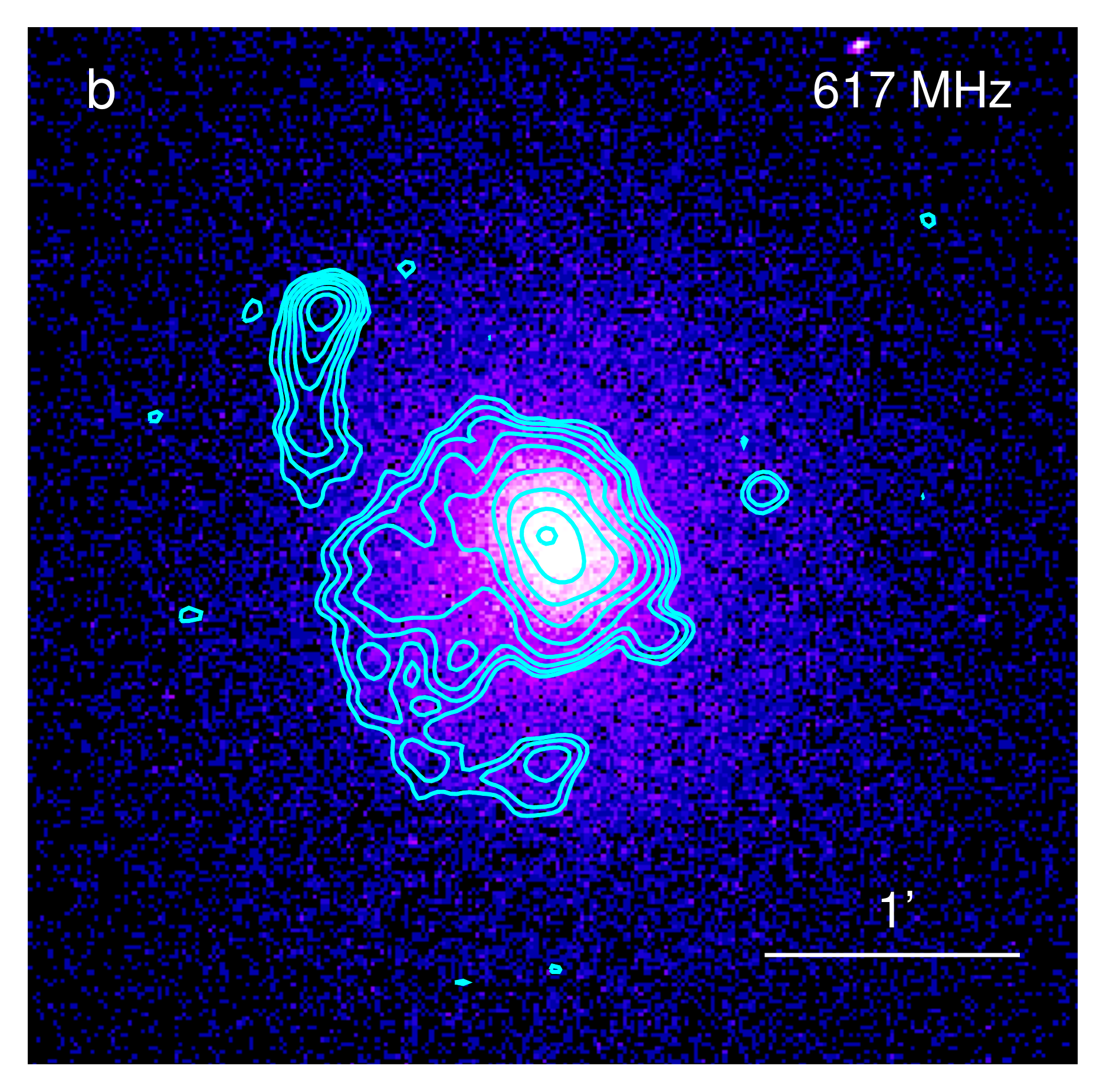}
\includegraphics[width=7.8cm]{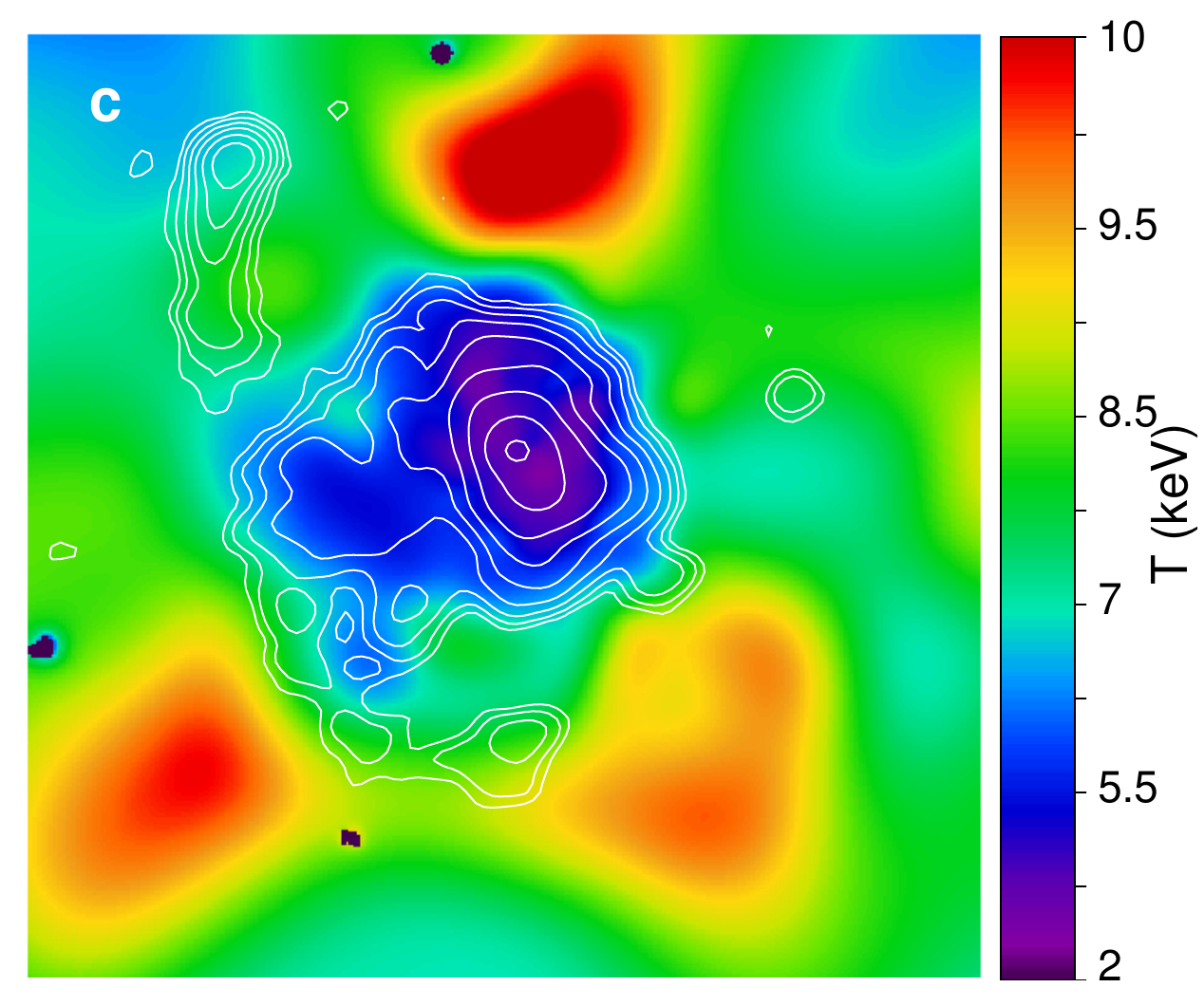}
\includegraphics[width=8.5cm]{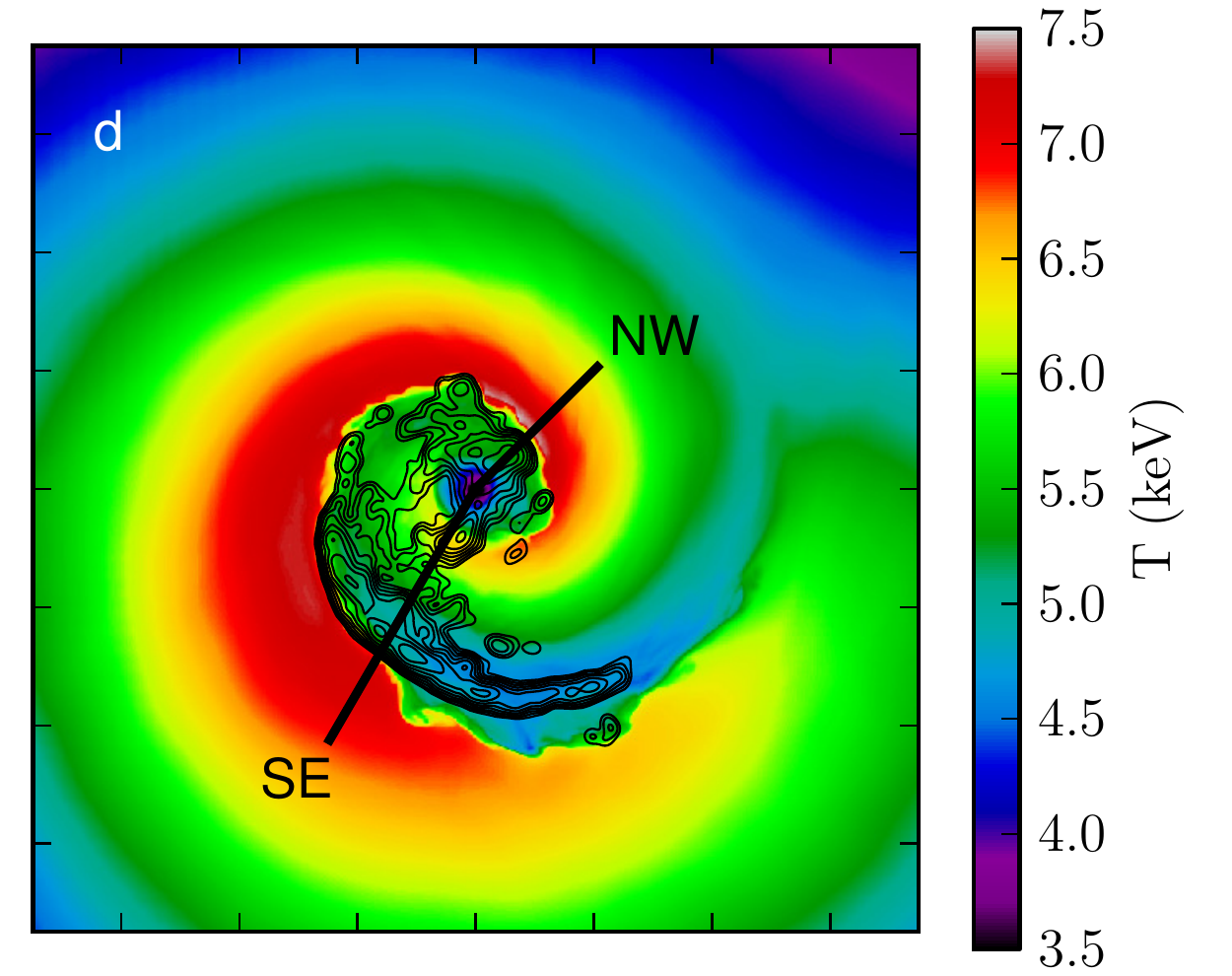}
\smallskip

\caption{(a) {\em Chandra} X-ray image of RX\,J1720.1 in
the 0.5-2.5 keV energy band. The background has been subtracted 
and the image has been divided by the exposure map and binned 
by 4 pixels. The dashed lines show 
the sectors used to extract the radial profiles shown in 
Fig.~\ref{fig:profile}. (b) Same {\em Chandra} image, with the 
\gmrt\ 617 MHz contours overlaid (from Fig.~2a). The X-ray cold fronts bound 
the minihalo emission, whose tail traces the SE cold front.  
(c) {\em Chandra} projected temperature map, 
with the 617 MHz contours overlaid (same as in (b)). 
(d) Contours of synchrotron emission at 327 MHz from the
turbulent reacceleration-model simulation, overlaid on the projected gas 
temperature map in the $z$-projection (from Z13). Contours 
increase by a factor of 2 starting from $0.5\times10^{-3}$~mJy arcsec$^{-2}$. 
The panel is 750~kpc on a side. Tick marks indicate 100~kpc distances.} 
\label{fig:mh_chandra}
\end{figure*}

In Fig.~\ref{fig:mh_chandra}(a), we show a {\em Chandra}\/ X-ray image of
RX\,J1720.1, obtained from the combination of three observations
(ObsIDs 1453, 3224 and 4631, for a total clean exposure of 42.5 ks; 
see Mazzotta \& Giacintucci 2008 for details),
showing the complex core of this otherwise relaxed cluster
(Fig.~\ref{fig:field}). Two cold fronts, located on the opposite sides from
the cluster center, appear to form a spiral structure that is seen in
numerous simulations of sloshing of the central low-entropy gas in
cluster cores (e.g., Ascasibar \& Markevitch 2006, Zuhone et al.\ 2011).  In
panel (b), we overlay the 617 MHz radio brightness contours of the minihalo
on the same X-ray image. As previously noticed by Mazzotta \& Giacintucci
(2008), the radio emission appears entirely contained within these cold
fronts. The new, higher-sensitivity radio image shows that the minihalo tail
is more extended than it was in the earlier data, and traces the SE cold
front remarkably well.

In panel (c), we present an overlay of the radio contours on the {\em Chandra}\/ projected 
temperature map, obtained using the observations ObsID 3224 and 4631 
(for a total clean exposure of 34.5 ks) following the algorithm described in \cite{2008A&A...479..307B}. 
Temperature values are derived from spectra from
overlapping square bins of varying scales, allowing us to map the
temperature variations using a B2-spline wavelet transform. This algorithm
has been adapted to the {\em Chandra}\/ ACIS-I instrument responses, using
the background model of Bartalucci et al.\ (2014). The wavelet transform has
been thresholded at $1\sigma$ and detects significant features on angular
scales $0.5^{\prime\prime}-8^{\prime\prime}$. The radio emission correlates
well with the cool gas spiral structure seen in the core of RX\,J1720.1. 
Panel (d) shows a snapshot from Z13 simulations of a radio minihalo 
in a relaxed cluster of similar mass, formed by turbulent
reacceleration of electrons in a sloshing cool core. 
The similarity of
simulations with the minihalo in RX\,J1720.1 is striking.

\begin{figure*}
\centering
\includegraphics[width=8cm]{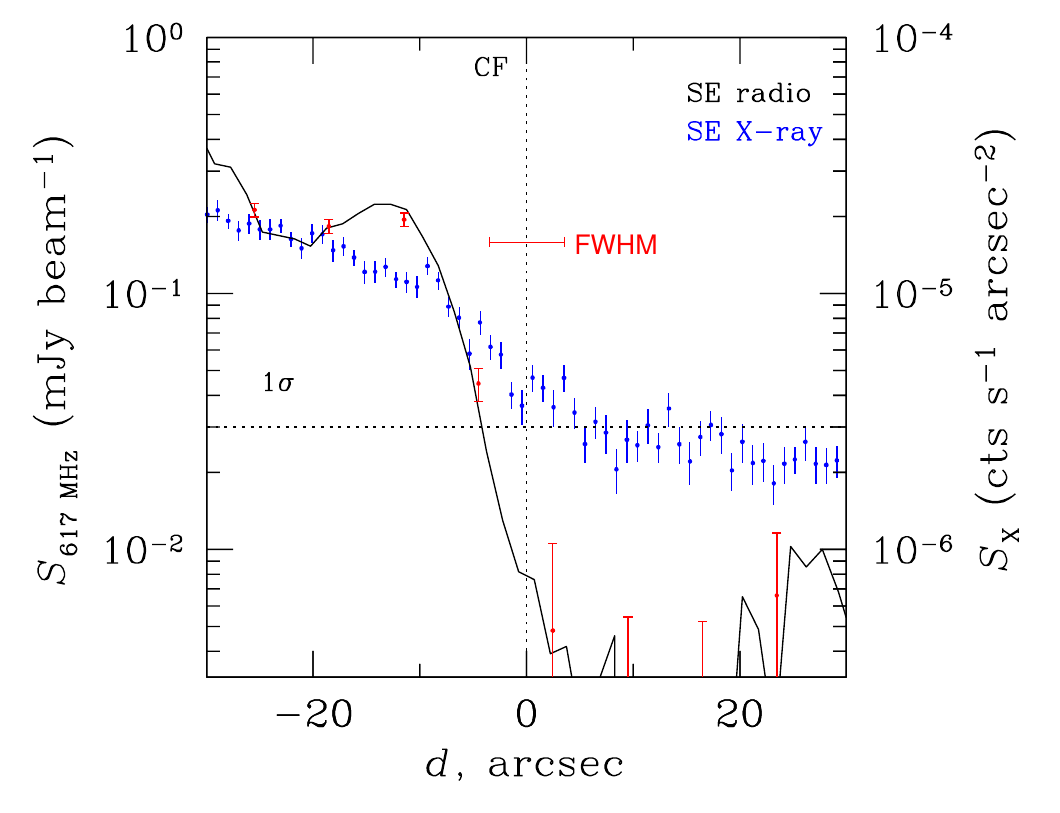}
\includegraphics[width=8cm]{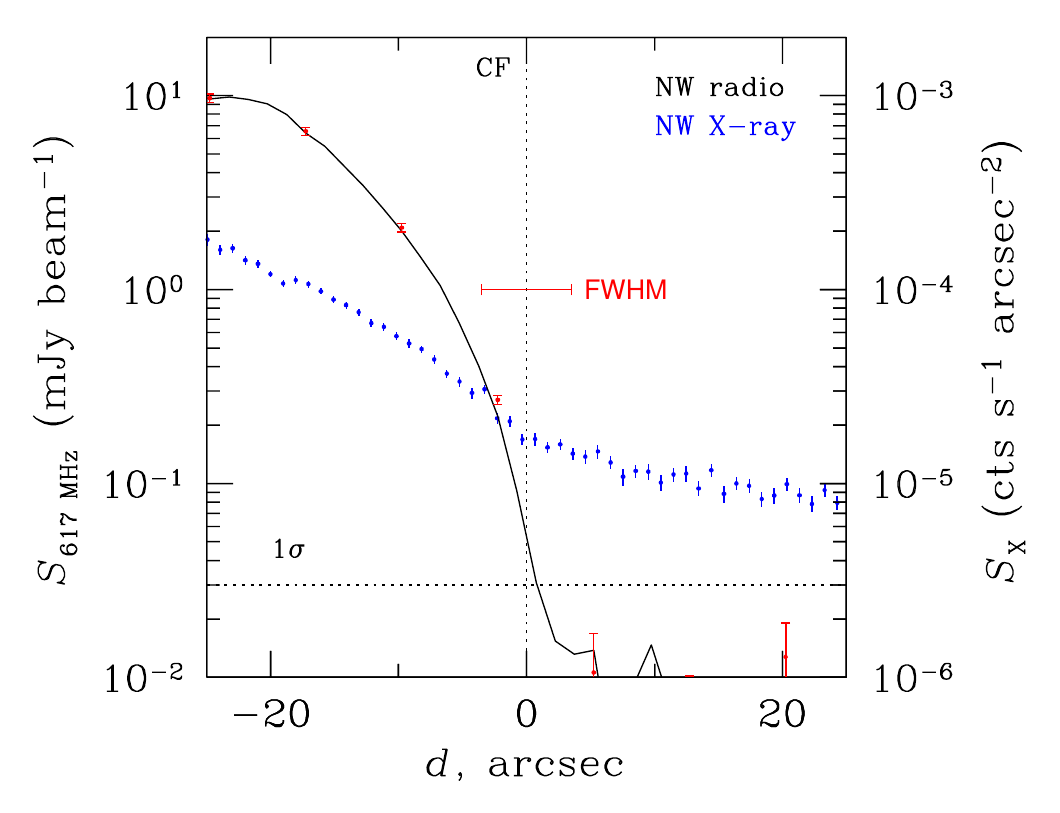}
\vspace{-0.5cm}
\includegraphics[width=8cm]{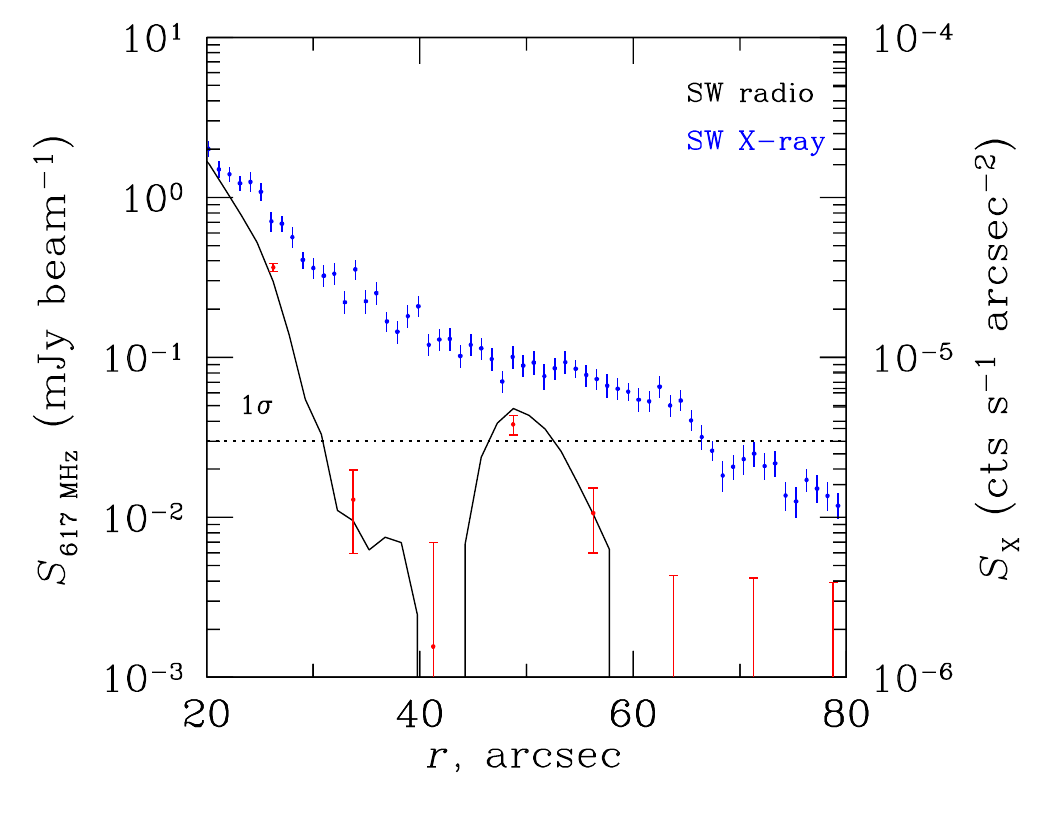}
\includegraphics[width=8cm]{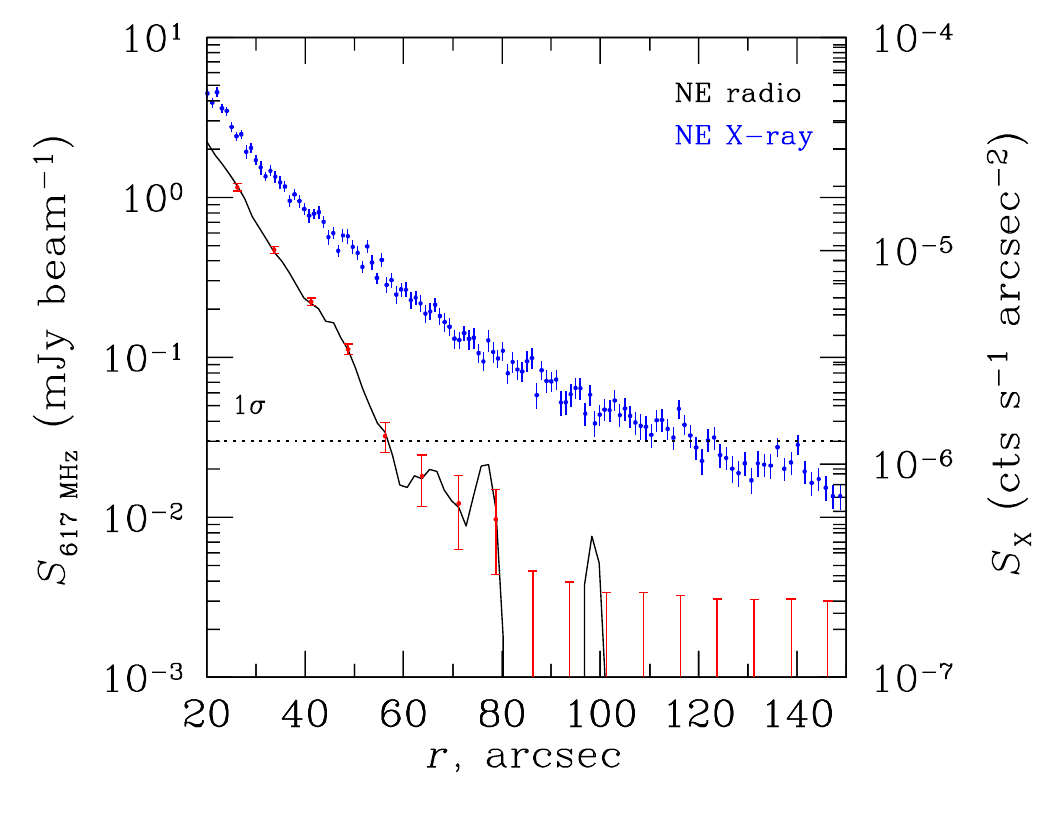}
\vspace{-0.5cm}
\includegraphics[width=7.2cm]{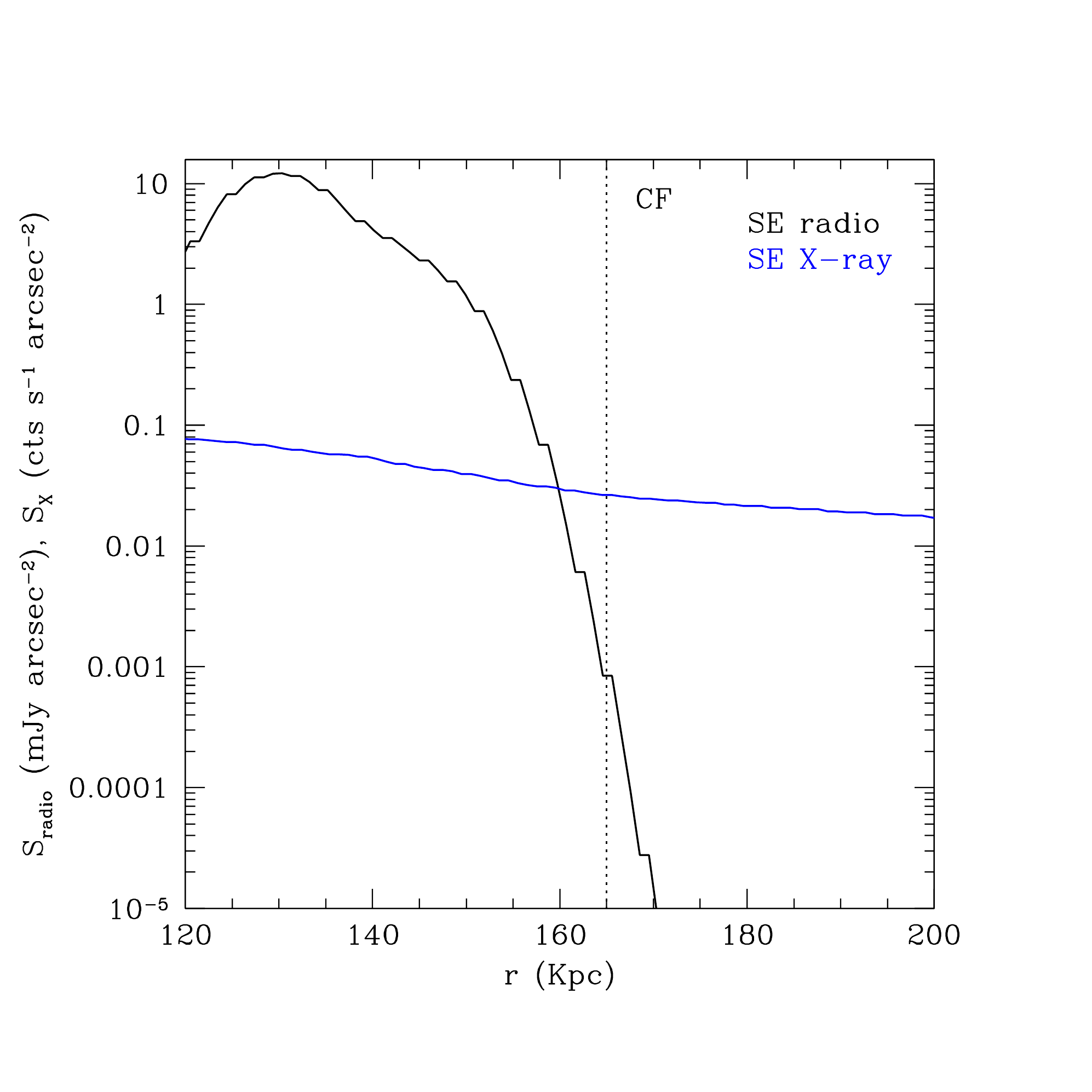}
\hspace{0.6cm}
\includegraphics[width=7.2cm]{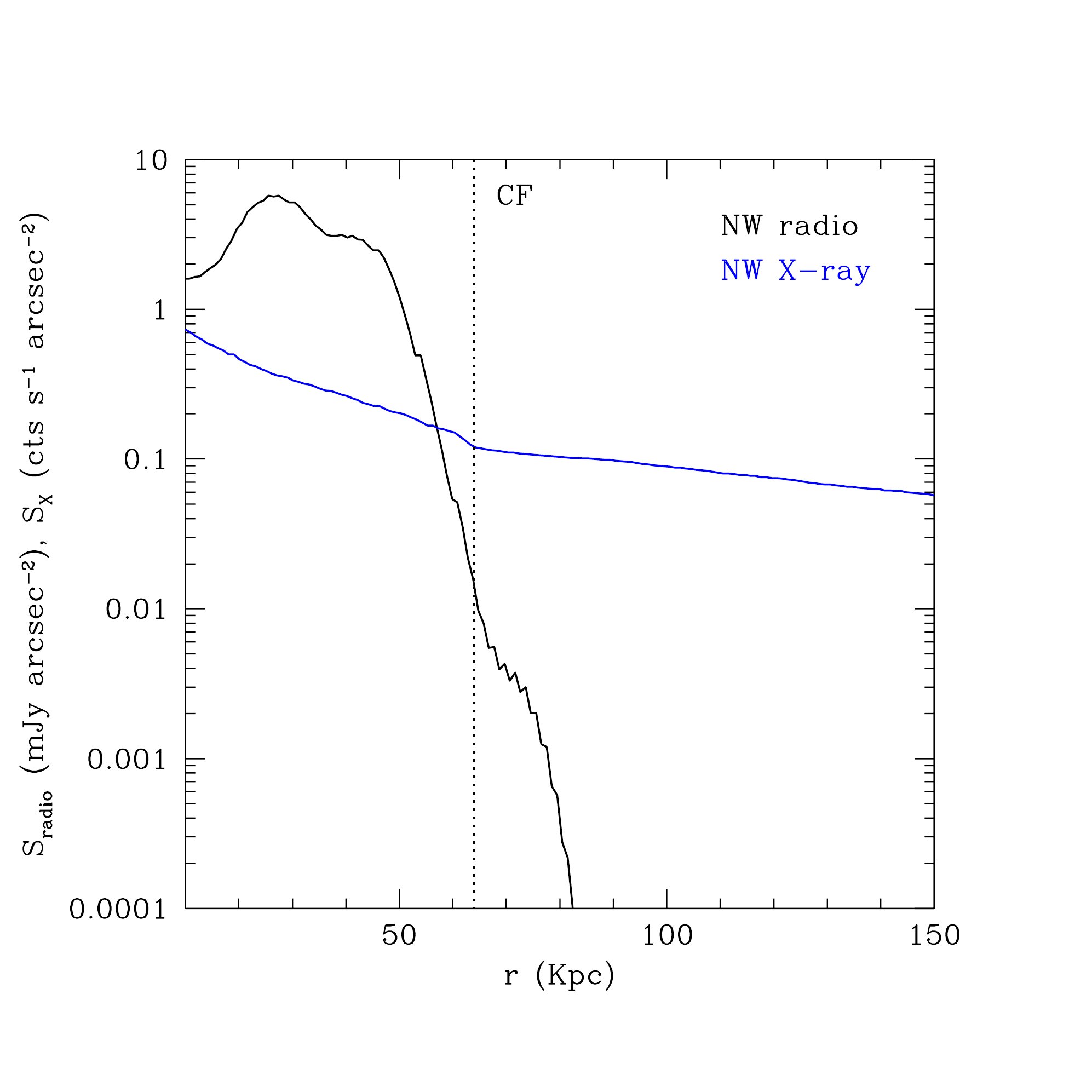}
\caption{{\em Upper panels:} X-ray (blue) and 617 MHz (black and red) 
brightness profiles, extracted in the sectors containing the X-ray cold fronts 
indicated in Fig.~\ref{fig:mh_chandra}(a). The x-axis has zero at the cold 
front radii (vertical, black-dotted lines). Error bars are $1\sigma$. The black 
profiles has been obtained using radial steps of $1^{\prime\prime}.5$. The 
red points are from steps as wide as the $7.5^{\prime\prime}$ FWHM 
(also shown in the plots). The horizontal, black-dotted lines indicates 
the $1\sigma$ noise level of the radio image. {\em Middle panels:}
Radial X-ray and radio profiles (as in the upper panels), 
extracted in $90^{\circ}$ SW and NE quadrants centered on the X-ray peak. 
{\em Bottom panels:} Profiles of the X-ray and radio emission  
from the reacceleration-model simulation across the SE and NW cold front 
surfaces (Fig.~\ref{fig:mh_chandra}(d)). The position of the 
cold fronts are marked with black-dotted lines. The emission units in the bottom panels 
have been renormalized so that all profiles are unity at $r=0$.}
\label{fig:profile}
\end{figure*}

The radial profiles of the radio and X-ray brightness in the upper panels of
Fig.~\ref{fig:profile} show the confinement of the minihalo within the cold
fronts more clearly. The profiles were extracted in the NW and SE sectors
shown by dashes in Fig.~\ref{fig:mh_chandra}(a). The profiles were centered
on the center of curvature of the fronts (RA= 17h 20m 10.3s, Dec=$+26^{\circ}
\, 37^{\prime} \, 20^{\prime\prime}$ for SE and RA=17h 20m 11.6s,
Dec=$+26^{\circ} \, 37^{\prime} \, 19^{\prime\prime}$ for NW, respectively).
The $x$-axis has a zero at the X-ray cold front radii ($r=57^{\prime\prime}$
and $r=51^{\prime\prime}$, respectively). 

At cold fronts, the X-ray brightness shows the subtle edges typical of
sloshing cold fronts. However, the radio profiles show an abrupt drop at
those positions --- no significant radio emission is seen beyond the fronts,
even with the high sensitivity of our data that allows us to detect
radio brightness at least two orders of magnitude below the peak of the minihalo. 
We stress that this behavior of the radio brightness is not an artifact 
of the radio image reconstruction
 from the interferometric data, since the largest detectable angular 
scale at 617 MHz (Table 2) is almost an order of magnitude larger 
than the size of the minihalo. 

In the middle panels of Fig.~\ref{fig:profile}, we show the radial profiles 
extracted in 90$^{\circ}$ NE and SW quadrants 
centered on the cluster X-ray peak, where no X-ray cold fronts are visible. 
Unlike the smooth X-ray brightness profiles, the radio brightness decreases 
sharply in those directions as well. 

The observed behavior of the radio and X-ray profiles 
is similar to that seen in simulations that assume the 
turbulent reacceleration origin of the radio emission, as shown in the 
lower panels of Fig.~\ref{fig:profile} (Z13).
The radio brightness in the simulated minihalo cuts off exactly 
at the positions of the X-ray cold fronts. The reason is that turbulence 
is confined to the volume enclosed by the cold fronts.

%

%

\subsection{Source of seed relativistic electrons}

Minihalos usually surround a compact radio source associated with the
central AGN, whose powerful explosions blow the radio bubbles commonly seen
filling the cavities in the X-ray gas (e.g., Perseus, Fabian et al. 2011;
McNamara \& Nulsen 2007 for a review). These AGNs cannot be the primary
source of the radio-emitting electrons in minihalos, because the time
required for the electrons to spread around the minihalo is too long
compared to their radiative cooling time (e.g., Jaffe 1977; see our
estimates for diffusion and advection in \S\ref{sec:origin} for the minihalo
in RXJ1720.1). However, the central AGN can be the source of seed electrons
for in-situ reacceleration \citep[e.g.,][]{2008A&A...486L..31C}. Large-scale
sloshing motions may disrupt the old radio bubbles inflated by AGN outbursts
and redistribute the aging relativistic plasma throughout the core region,
where turbulence then picks them up and reaccelerates (e.g., Z13; see also
discussion in Giacintucci et al.\ 2014). As mentioned above, hadronic
collisions, which should produce relativistic electrons throughout the
cluster volume, may also be the source of seed electrons.

In RX\,J1720.1, the central radio galaxy is weak ($\sim 5\times
10^{23}$ W Hz$^{-1}$ at 1.48 GHz) and compact ($<1.4$ kpc). No X-ray
cavities are visibile in the cluster core region, consistent with the
absence of any radio lobes and jets on the corresponding $\sim 10$ kpc
scales.  However, unlike the flat spectra of core-dominated radio galaxies
and sources in which the emission is dominated by the beamed radiation of a
jet pointing toward us \citep[e.g.,][]{2013ApJS..208...15M}, the spectrum of the
RXJ1720.1 radio galaxy is relatively steep ($\alpha\sim 0.8$) and similar to
that of extended, active radio galaxies. This suggests that the AGN is in
the active stage and possesses radio jets/lobes, but they are on sub-kpc
scales. They could be revealed by radio observations with higher angular
resolution. It is possible that the radio galaxy is small because it is
young --- perhaps it restarted recently after the cessation of the previous
cycle of nuclear activity. The ongoing sloshing motion of the gas core may
have had enough time to break up the old, faded radio lobes from the past
cycles, and spread their relativistic content throughout the sloshing
region.  Interestingly, Z13, in their simulation of a sloshing core,
considered an initial distribution of seed relativistic electrons in the
shape of two filled spheres, mimicking two radio bubbles. Sloshing mixed
them thoroughly, and the resulting radio map was very similar to that 
from the alternative model with the initial uniform distribution of the seed
electrons --- the final distribution of the radio brightness was determined
by the locations of the strongest gas turbulence.

\section{Summary and conclusions}
\label{sec:summ}

We presented multi-frequency GMRT and VLA observations of the radio minihalo
in the cool core of RX\,J1720.1, which constitute the most detailed
radio dataset for this class of objects to date. The RXJ1720.1 minihalo
consists of a bright central region that contains most of its flux density,
and a $\sim 230$ kpc-long, arc-shaped tail of lower surface brightness.

Based on our flux density measurements at six frequencies between 317 MHz
and 8.44 GHz, we studied the integrated radio spectrum of the minihalo and
its components. We found indication of a possible steepening of the 
spectrum above 5 GHz. This steepening is seen separately in the spectrum 
of the central region of the minihalo and possibly in the spectrum of the tail. 
Deeper, high-frequency observations are necessary to confirm this 
steepening and quantify the change of the spectral slope. If confirmed,
the presence of a break has 
important implications for the physical mechanism responsible for the radio-emitting 
relativistic electrons.

The map of the spectral index between 617 MHz and 1.48 GHz shows that the
spectrum steepens systematically with increasing distance from the center,
in particular, along the tail of the minihalo. We have shown that
interpretations of this steepening that involve diffusion or advection of
electrons produced in the central region and their aging along the way
require extreme microphysical conditions or implausible gas velocities.
The proposed mechanisms for the origin of minihalos are turbulent reacceleration 
and continuous injection of secondary electrons due to inelastic collisions between
 relativistic and thermal protons. The presence of a possible spectral break
and strong spatial variations of the spectral index challenge the ``secondary'' 
origin of the minihalo and favors reacceleration by turbulence. As shown 
in MHD simulations (Z13), the 
required turbulence can be generated by sloshing of the 
low-entropy gas in the cluster cool core. Sloshing also amplifies 
the magnetic field in the core, increasing the radio emission from the
cosmic-ray electrons. The turbulence should be limited to the 
volume enclosed by cold fronts visible in the X-ray. This scenario produces 
a radio minihalo entirely
contained within the sharp boundaries at the positions of cold fronts ---
exactly as we observe in RXJ1720.1.
\\
\\
{\it Acknowledgements.}

The authors thank the anonymous referee, whose comments and suggestions 
improved the paper. SG thanks Tracy Clarke for useful discussions.
SG acknowledges the support of NASA through Einstein Postdoctoral
Fellowship PF0-110071 awarded by the Chandra X-ray Center (CXC), which
is operated by SAO. JAZ is supported under the NASA Postdoctoral Program.
GMRT is run by the National Centre for Radio Astrophysics of the Tata Institute 
of Fundamental Research. The National Radio Astronomy Observatory is a facility 
of the National Science Foundation operated under cooperative agreement by 
Associated Universities, Inc. The scientific results reported in this article are based 
on observations made by the {\em Chandra} X-ray Observatory.

{}


\begin{thebibliography}{}




\bibitem[Ascasibar \& Markevitch(2006)]{2006ApJ...650..102A} Ascasibar, Y., \& Markevitch, M.\ 2006, \apj, 650, 102 

\bibitem[Bartalucci et al.(2014)]{} Bartalucci, I., Mazzotta, P., Bourdin, H., Vikhlinin, A., 2014, A\&A, submitted 

\bibitem[Berezinskii et al.(1990)]{1990acr..book.....B} Berezinskii, V.~S., 
Bulanov, S.~V., Dogiel, V.~A., 
\& Ptuskin, V.~S.\ 1990, Amsterdam: North-Holland, 1990, edited by Ginzburg, V.L.,  




\bibitem[B{\"o}hringer et al.(2000)]{2000ApJS..129..435B} B{\"o}hringer, 
H., Voges, W., Huchra, J.~P., et al.\ 2000, \apjs, 129, 435 

\bibitem[Bourdin 
\& Mazzotta(2008)]{2008A&A...479..307B} Bourdin, H., \& Mazzotta, P.\ 2008, \aap
, 479, 307 

\bibitem[Brunetti et al.(2007)]{2007ApJ...670L...5B} Brunetti, G., Venturi, 
T., Dallacasa, D., et al.\ 2007, \apjl, 670, L5 

\bibitem[Brunetti 
\& Jones(2014)]{2014IJMPD..2330007B} Brunetti, G., \& Jones, T.~W.\ 2014, International Journal of Modern Physics D, 23, 30007 


\bibitem[Cassano et 
al.(2008)]{2008A&A...486L..31C} Cassano, R., Gitti, M., \& Brunetti, G.\ 2008, \aap, 486, L31 

\bibitem[Cassano et al.(2013)]{2013ApJ...777..141C} Cassano, R., Ettori, 
S., Brunetti, G., et al.\ 2013, \apj, 777, 141 

\bibitem[Cavagnolo et al.(2009)]{2009ApJS..182...12C} Cavagnolo, K.~W., 
Donahue, M., Voit, G.~M., \& Sun, M.\ 2009, \apjs, 182, 12 

\bibitem[Chandra et al.(2004)]{2004ApJ...612..974C} Chandra, P., Ray, A., 
\& Bhatnagar, S.\ 2004, \apj, 612, 974 



\bibitem[Condon(1992)]{1992ARA&A..30..575C} Condon, J.~J.\ 1992, \araa, 30, 575 



\bibitem[Dupke 
\& White(2003)]{2003ApJ...583L..13D} Dupke, R., \& White, R.~E., III 2003, \apjl, 583, L13 


\bibitem[Ettori et 
al.(2013)]{2013A&A...555A..93E} Ettori, S., Gastaldello, F., Gitti, M., et al.\ 2013, \aap, 555, A93 

\bibitem[Fabian et al.(2011)]{2011MNRAS.418.2154F} Fabian, A.~C., Sanders, 
J.~S., Allen, S.~W., et al.\ 2011, \mnras, 418, 2154 

\bibitem[Feretti 
\& Venturi(2002)]{2002ASSL..272..163F} Feretti, L., \& Venturi, T.\ 2002, Merging Processes in Galaxy Clusters, 272, 163 


\bibitem[Fujita et al.(2004)]{2004ApJ...612L...9F} Fujita, Y., Matsumoto, 
T., \& Wada, K.\ 2004, \apjl, 612, L9 

\bibitem[Fujita et al.(2007)]{2007ApJ...663L..61F} Fujita, Y., Kohri, K., 
Yamazaki, R., \& Kino, M.\ 2007, \apjl, 663, L61 

\bibitem[Fujita 
\& Ohira(2012)]{2012ApJ...746...53F} Fujita, Y., \& Ohira, Y.\ 2012, \apj, 746, 53

\bibitem[Fujita 
\& Ohira(2013)]{2013MNRAS.428..599F} Fujita, Y., \& Ohira, Y.\ 2013, \mnras, 428, 599 

\bibitem[Ghizzardi et 
al.(2010)]{2010A&A...516A..32G} Ghizzardi, S., Rossetti, M., \& Molendi, S.\ 2010, \aap, 516, A32 

\bibitem[Giacintucci et 
al.(2008)]{2008A&A...486..347G} Giacintucci, S., Venturi, T., Macario, G., et al.\ 2008, \aap, 486, 347 

\bibitem[Giacintucci et al.(2011)]{2011ApJ...732...95G} Giacintucci, S., 
O'Sullivan, E., Vrtilek, J., et al.\ 2011, \apj, 732, 95

\bibitem[Giacintucci et al.(2014)]{2014ApJ...781....9G} Giacintucci, S., 
Markevitch, M., Venturi, T., et al.\ 2014, \apj, 781, 9 

\bibitem[Gitti et 
al.(2002)]{2002A&A...386..456G} Gitti, M., Brunetti, G., \& Setti, G.\ 2002, \aap, 386, 456 

\bibitem[Hess et al.(2012)]{2012AJ....144...48H} Hess, K.~M., Wilcots, 
E.~M., \& Hartwick, V.~L.\ 2012, \aj, 144, 48 


\bibitem[Hlavacek-Larrondo et al.(2011)]{2011MNRAS.415.3520H} 
Hlavacek-Larrondo, J., Fabian, A.~C., Sanders, J.~S., 
\& Taylor, G.~B.\ 2011, \mnras, 415, 3520 

\bibitem[Hlavacek-Larrondo et al.(2013)]{2013ApJ...777..163H} 
Hlavacek-Larrondo, J., Allen, S.~W., Taylor, G.~B., et al.\ 2013, \apj, 
777, 163 

\bibitem[Jaffe(1977)]{1977ApJ...212....1J} Jaffe, W.~J.\ 1977, \apj, 212, 1 

\bibitem[Kale al.(2013)]{2013A&A...557A..99K} Kale, R., Venturi, T., Giacintucci, S., et al.\ 2013, \aap, 557, A99 

\bibitem[Keshet(2010)]{2010arXiv1011.0729K} Keshet, U.\ 2010, 
arXiv:1011.0729 


\bibitem[Keshet 
\& Loeb(2010)]{2010ApJ...722..737K} Keshet, U., \& Loeb, A.\, 2010, \apj, 722, 737 

\bibitem[Keshet et al.(2010)]{2010ApJ...719L..74K} Keshet, U., Markevitch, 
M., Birnboim, Y., \& Loeb, A.\ 2010, \apjl, 719, L74 

\bibitem[Kuchar 
\& En{\ss}lin(2011)]{2011A&A...529A..13K} Kuchar, P., \& En{\ss}lin, T.~A.\ 2011, \aap, 529, A13 


\bibitem[Lane et al.(2012)]{2012RaSc...47.0K04L} Lane, W.~M., Cotton, 
W.~D., Helmboldt, J.~F., \& Kassim, N.~E.\ 2012, Radio Science, 47, 0 


\bibitem[Markevitch et al.(2001)]{2001ApJ...562L.153M} Markevitch, M., 
Vikhlinin, A., \& Mazzotta, P.\ 2001, \apjl, 562, L153 


\bibitem[Markevitch \& Vikhlinin(2007)]{2007PhR...443....1M} Markevitch, M., \& Vikhlinin, A.\ 2007, \physrep, 443, 1 

\bibitem[Massaro et al.(2013)]{2013ApJS..208...15M} Massaro, F., Giroletti, 
M., Paggi, A., et al.\ 2013, \apjs, 208, 15 


\bibitem[Mazzotta et al.(2001)]{2001ApJ...555..205M} Mazzotta, P., 
Markevitch, M., Vikhlinin, A., et al.\ 2001, \apj, 555, 205 

\bibitem[Mazzotta et al.(2003)]{2003ApJ...596..190M} Mazzotta, P., Edge, 
A.~C., \& Markevitch, M.\ 2003, \apj, 596, 190 

\bibitem[Mazzotta 
\& Giacintucci(2008)]{2008ApJ...675L...9M} Mazzotta, P., \& Giacintucci, S.\ 2008, \apjl, 675, L9 


\bibitem[Mazzotta 
\& Giacintucci(2008)]{2008ApJ...675L...9M} Mazzotta, P., \& Giacintucci, S.\ 2008, \apjl, 675, L9 

\bibitem[McNamara 
\& Nulsen(2007)]{2007ARA&A..45..117M} McNamara, B.~R., \& Nulsen, P.~E.~J.\ 2007, \araa, 45, 117 


\bibitem[Murgia et 
al.(2010)]{2010A&A...514A..76M} Murgia, M., Eckert, D., Govoni, F., et al.\ 2010, \aap, 514, A76 

\bibitem[Owers et al.(2009)]{2009ApJ...704.1349O} Owers, M.~S., Nulsen, 
P.~E.~J., Couch, W.~J., \& Markevitch, M.\ 2009, \apj, 704, 1349 

\bibitem[Owers et al.(2011)]{2011ApJ...741..122O} Owers, M.~S., Nulsen, 
P.~E.~J., \& Couch, W.~J.\ 2011, \apj, 741, 122 

\bibitem[Perley 
\& Butler(2013)]{2013ApJS..204...19P} Perley, R.~A., \& Butler, B.~J.\ 2013, \apjs, 204, 19 



\bibitem[Pfrommer 
\& En{\ss}lin(2004)]{2004A&A...413...17P} Pfrommer, C., \& En{\ss}lin, T.~A.\ 2004, \aap, 413, 17 

\bibitem[Piffaretti et 
al.(2011)]{2011A&A...534A.109P} Piffaretti, R., Arnaud, M., Pratt, G.~W., Pointecouteau, E., \& Melin, J.-B.\ 2011, \aap, 534, A109 

\bibitem[Planck Collaboration et al.(2013)]{2013arXiv1303.5089P} Planck 
Collaboration, Ade, P.~A.~R., Aghanim, N., et al.\ 2013, arXiv:1303.5089 


\bibitem[Roediger et al.(2011)]{2011MNRAS.413.2057R} Roediger, E., 
Br{\"u}ggen, M., Simionescu, A., et al.\ 2011, \mnras, 413, 2057 


\bibitem[Scaife 
\& Heald(2012)]{2012MNRAS.423L..30S} Scaife, A.~M.~M., \& Heald, G.~H.\ 2012, \mnras, 423, L30 

\bibitem[Sijbring (1993)]{} Sijbring L.G., PhD Thesis, University of Groningen, A radio continuum and HI Line study of the Perseus cluster (1993)

\bibitem[Tittley 
\& Henriksen(2005)]{2005ApJ...618..227T} Tittley, E.~R., \& Henriksen, M.\ 2005, \apj, 618, 227 

\bibitem[Vazza et 
al.(2012)]{2012A&A...544A.103V} Vazza, F., Roediger, E., \& Br{\"u}ggen, M.\ 2012, \aap, 544, A103 

\bibitem[Venturi et 
al.(2008)]{2008A&A...484..327V} Venturi, T., Giacintucci, S.,
Dallacasa, D., et al.\ 2008, \aap, 484, 327 


\bibitem[Zandanel et al.(2014)]{2014MNRAS.438..124Z} Zandanel, F., 
Pfrommer, C., \& Prada, F.\ 2014, \mnras, 438, 124 

\bibitem[ZuHone et al.(2011)]{2011ApJ...743...16Z} ZuHone, J.~A., 
Markevitch, M., \& Lee, D.\ 2011, \apj, 743, 16 



\bibitem[ZuHone et al.(2013)]{2013ApJ...762...78Z} ZuHone, J.~A., 
Markevitch, M., Brunetti, G., \& Giacintucci, S.\ 2013, \apj, 762, 78 (Z13)

\bibitem[ZuHone et al.(2014)]{} ZuHone, J.~A.,  Brunetti, G., Giacintucci, S., Markevitch, M., 2014,
\apj, submitted, arXiv:1403.6743


\end{thebibliography}
\end{document}